\documentclass[twocolumn]{aa} % for a referee version
\usepackage{graphicx}
\usepackage{subfigure}
\usepackage{natbib}
\usepackage{txfonts}
\begin{document}
   \title{On the influence of collisional rate coefficients\\ on the water vapour excitation}

%   \subtitle{I. Overviewing the $\kappa$-mechanism}

\titlerunning{H$_2$O excitation}
\authorrunning{Daniel et al.}
\author{F. Daniel\inst{1}, J.R. Goicoechea\inst{1}, J. Cernicharo\inst{1}, M.-L. Dubernet\inst{2,3}, A. Faure \inst{4}}
\institute{Departamento de Astrof\'isica, Centro de
Astrobiolog\'ia, CSIC-INTA, Ctra. de Torrej\'on a Ajalvir km 4,
28850 Madrid, Spain; \email{danielf@cab.inta-csic.es} 
\and 
Observatoire de Paris, LUTH UMR CNRS 8102, 5 Place Janssen, 92195 Meudon, France
\and
Universit\'e Pierre et Marie Curie, LPMAA UMR CNRS 7092, Case 76,
4 Place Jussieu, 75252 Paris Cedex 05, France
\and
UJF-Grenoble 1/CNRS-INSU, Institut de Plan\'etologie et d'Astrophysique de Grenoble (IPAG) UMR 5274, Grenoble F-38041, France}

   \date{Received ...; accepted ...}

% \abstract{}{}{}{}{} 
% 5 {} token are mandatory
 
  \abstract
   {Water is a key molecule in many astrophysical studies that deal with star and planet forming regions, evolved stars and galaxies.
   Its high dipole moment makes this molecule to be subthermally populated under the typical conditions of most
   astrophysical objects. This motivated the calculation of
    various sets of collisional rate coefficients (CRC) for H$_2$O (with He or H$_2$) which are 
    necessary to model its rotational excitation and line emission.}
   {The most accurate set of CRC are the quantum rates that involve H$_2$.
   However, they were published only recently and less accurate CRC
   (quantum with He or quantum classical trajectory (QCT) with H$_2$) were used in many studies. This work aims to underline 
   the impact that the new available set of CRC have on the interpretation of water vapour observations.}
   {We performed accurate non--local non--LTE radiative transfer calculations using different sets of CRC in order 
   to predict the line intensities from transitions that involve the lowest energy levels of H$_2$O (E $<$ 900 K). The results obtained from
   the different CRC sets are then compared using line intensity ratio statistics.}
   {For the whole range of physical conditions considered in this work, we obtain that the intensities based on the quantum and 
   QCT CRC are in good agreement.
   However, at relatively low H$_2$ volume density ($n$(H$_2$) $<$ 10$^7$ cm$^{-3}$) 
   and low water abundance ($\chi$(H$_2$O) $<$ 10$^{-6}$), these physical conditions being relevant to describe
   most molecular clouds, we find differences in the predicted line intensities of up to a factor of $\sim$ 3 for the bulk of the lines.        
   Most of the recent studies interpreting early Herschel Space Observatory spectra used the QCT CRC.
   Our results show that although the global conclusions from those studies will not be drastically changed, 
   each case has to be considered individually, since depending on the physical conditions, the use of the QCT CRC may lead 
   to a mis--estimate of the water vapour abundance of up to a factor of $\sim$ 3.
   Additionally, the comparison of the quantum state--to--state
   and thermalized CRC that include the description of the population of the H$_2$ rotational levels
   show that above T$_K$ $\sim$ 100 K, large differences are expected from those two sets for the p--H$_2$ symmetry.
   Finally, we have found that at low temperature (i.e. T$_K$ $<$ 100 K) modelled line intensities 
   will be differentially affected by the symmetry of the H$_2$ molecule. 
   If a significant number of H$_2$O lines is observed, it is therefore possible to obtain an estimate of the H$_2$ 
   ortho--to--para ratio from the analysis of the line intensities.}
   {}


   \maketitle
%
%________________________________________________________________

\section{Introduction}

Water is a key molecule for both the chemistry and the cooling budget along the star formation trail.
The determination of the water vapour abundance is
a long standing problem in astrophysics. Because water
vapour is predicted to be an abundant molecule in the
gas phase, the determination of its spatial extent, its
distribution, and its abundance is crucial in modelling
the chemistry and the physics of molecular clouds, comets, evolved
stars and galaxies \citep{Neu93, Neufeld95,
cernicharo2005, vandishoek2011}. In warm molecular clouds, water vapour can play a
critical role in the gas cooling \citep{Neu93, Neufeld95}
and, hence, in the evolution of these objects.

Unfortunately, water is an abundant molecule in our  atmosphere making particularly 
difficult the observation  of its rotational lines and vibrational bands
from Earth. Even so, some 
observations of H$_2$O maser lines have been performed from ground--based
and airborne telescopes: the $6_{16}-5_{23}$  at
22~GHz \citep{Che69}, the $5_{15}-4_{22}$  at
325~GHz \citep{Men90a} , the $10_{29}-9_{36}$ 
at 321~GHz \citep{Men90b}  and the $3_{13}-2_{20}$ 
at 183.31~GHz \citep{Wat80, Cer90, Cer94}.
The spectrometers on board the \textit{Infrared Space Observatory} \citep{Kes96} 
provided a unique opportunity to observe infrared (IR\footnote{Throughout this article, many acronyms are used. While defined in the article when first introduced,  they are compiled here for clarity :
\begin{itemize}
\item IR: infrared
\item CRC: collisional rate coefficients
\item STS : state--to--state
\item RT: radiative transfer
\item QCT: quantum classical trajectory
\item OTPR : ortho--to--para ratio
\end{itemize}
}) H$_2$O thermal lines 
in a great variety of astronomical environments. 
Several studies demonstrated that water emission is a unique tracer of the warm gas 
and energetic processes taking place during star formation
(see reviews by \citet{vD04} and \citet{cernicharo2005}).
In particular, ISO observations showed that far-IR H$_2$O emission lines are 
important coolants of the warm gas affected by shocks (\textit{e.g.,} more than 70 pure rotational lines were
detected towards Orion~KL outflows; Cernicharo et al. 2006a) confirming earlier theoretical predictions of 
its importance in the shocked gas cooling \citep[e.g.,][]{Neu93}. 
ISO also detected widespread H$_2$O absorption towards the Galactic
Center \citep[e.g.,][]{Goi04}  and towards the nucleus of more distant 
galaxies  \citep[e.g.,][]{Gon04}. After ISO,
the launch of both the \textit{Submillimeter Wave Astronomy Satellite}, SWAS \citep{Mel00},
 and \textit{ODIN} \citep{Nor03}  allowed the
observation of the $1_{10}-1_{01}$ fundamental transition of both
H$_{2}^{16}$O (at 557~GHz)  and H$_{2}^{18}$O (at 548~GHz, first detected by the \textit{Kuiper Airborne
Observatory}; Zmuidzinas et al. 1995) at high heterodyne spectral resolution but poor angular resolution.
Their resolved line-profiles (line-wing emission, widths, self-absorption dips, etc.) were studied in detail. 
Finally, the \textit{Spitzer Space Telescope} has detected even higher excitation H$_2$O pure rotational lines
(up to $E_u$$\simeq$3000\,K) in the shocked gas around protostars (albeit at low spectral resolution;
\textit{e.g.,} Watson et al. 2007).

The HIFI and PACS spectrometers on board  \textit{Herschel Space Observatory}  \citep{Pil10}
provide much higher sensitivity and angular/resolution than previous far-IR observations,
allowing us  to detect a larger number of excited H$_2$O lines in many more sources,
 and to better constrain the spatial origin of the water vapour emission. 
The \textit{HIFISTARS} \footnote{http://hifistars.oan.es/}
key project has addressed the problem of the water abundance in evolved stars
and complement with high spectral resolution the results obtained previously by the Infrared Space
Observatory (ISO). Similar goals, but covering a much larger spectral domain with lower spectral
resolution, have been addressed by the \textit{MESS} (Mass--loss of Evolved StarS\footnote{http://www.univie.ac.at/space/MESS/}) key project. 
The \textit{WISH} (Water In Star forming regions with Herschel\footnote{http://www.strw.leidenuniv.nl/WISH/})  key project
focussed on the study of young stellar objects in different 
evolutionary stages. 
Early Herschel results include the detection of strong H$_2$O emission in protostellar environments
\citep[e.g.,][]{vandishoek2011}; the presence of water vapour  in
diffuse interstellar clouds with an \textit{ortho-to-para} ratio (OTPR)  consistent with the high temperature ratio of 3 
\citep{Lis10}, the detection of cold water vapour in TW Hydrae protoplanetary disk
with a low  OTPR of $\simeq$0.8 \citep{Hog11} and the widespread occurrence of H$_2$O in
circumstellar envelopes around  O-rich and C-rich evolved stars \citep{Roy10, Neu11}. 
In the outflows of Class~0 protostars, for example, tens of pure rotational lines
 of water vapour (up to $9_{18}-9_{09}$ or $E_u$$\simeq$1500\,K) are 
readily detected in the far-IR domain \citep{Her12, Goi12}.

In order to derive the water vapour abundance and to estimate the prevailing physical conditions in the
above environments, the energy level excitation and the radiative transfer (RT) of H$_2$O lines has 
to be understood. 
H$_2$O is an asymmetric molecule with a irregular set of energy levels characterised by
quantum numbers $J_{K_A\,K_C}$.
Because of the large rotation constants of H$_2$O, its pure rotational transitions lie in the 
submm and far-IR domain. Their high critical densities (much higher than CO lines) and  large optical-depths often 
results in a complex non-local and non-LTE excitation and RT problem.
In addition, in sources with strong far-IR continuum  emission, radiative pumping by warm dust photons can play
an important role in determining the rotational levels population \citep[e.g.,][]{Cer06b}.

Most of the information which is made available through water lines observations 
rely on modelling its excitation. From this point of view, water is difficult molecule
to treat since its high dipole moment makes most of its transitions to be subthermaly 
excited (see, e.g., \citet{Cer06b}), harbouring very large opacities (see, e.g.,
\citet{Gonzalez98}), many of them being maser in nature
\citep{Che69, Wat80, Phillips80, Men90a, Men90b, Menten91, Cer90, Cer94, Cernicharo96, Cernicharo99, Cer06b, 
Gonzalez95, Gonzalez98}. 
An accurate modelling thus require, in addition to a good description of the source structure, 
the availability of accurate collisional rate coefficients. In the case of saturated masers a special formalism has to
be developed in order to take into account saturation effects and to solve the RT problem \citep{Daniel12}.
To summarise, the water vapour abundance in different environments can change by orders of magnitude 
\citep[e.g.,][]{vandishoek2011} and H$_2$O rotational line profiles are sensitive probes of the gas 
kinematics and physical conditions \citep[e.g.,][]{Kri12}. These facts make water vapour lines a powerful
diagnostic tool in astrophysics. 

The methodology used to compare two collisional rate coefficients sets is presented in Sect. \ref{section:methodology}
and a comparison between the various sets available for H$_2$O is given in Sect. \ref{section:sets}.
A discussion on the effect introduced by the H$_2$ ortho--to--para ratio is presented in Sect. \ref{OTPR}.
Finally, we discuss the current results in Sect. \ref{discussion} and the conclusions are drawn in Sect. \ref{conclusions}.

%__________________________________________________________________

\section{Comparison between collisional rate coefficients sets: methodology} \label{section:methodology}

Water is a key molecule for both the chemistry and cooling budget 
of the warm molecular gas. The need to understand the excitation mechanisms
leading to line formation motivated the 
determination of various collisional 
rate coefficients sets (hereafter referred as CRC) for this molecule.

The first water vapour CRC were calculated 
using He as a collisional partner, considered
 the first $45^{th}$ rotational energy levels for both ortho-- and para--H$_2$O and 
 were calculated for temperatures in the range 20--2000 K \citep{green1993}. 
Collisions with H$_2$ were subsequently determined \citep{phillips1996} 
making use of the 5D potential energy surface (PES) described in \citet{phillips1994}.
This study showed
that considering either ortho-- or para--H$_2$ as a collisional 
partner could lead to substantial differences in the magnitude of the CRC. 
However, this study dealt with a limited number of H$_2$O energy 
levels (5 levels) and results were only made available for a reduced 
range of temperatures (20--140K). The latter calculations were 
extended to lower temperatures (to cover the range 5--20K) 
by \citet{dubernet2002} and \citet{grosjean2003} making use of the 
same PES. The importance of water subsequently leaded 
to the calculation of a high precision 9D PES for the H$_2$O -- H$_2$ 
system \citep{faure2005,valiron2008}. The influence of the results based on the latter 
PES (averaged over the vibration of H$_2$O and H$_2$ thus reducing the dimension to 5) 
with the previous PES \citep{phillips1994} are discussed in \cite{dubernet2006}.
Using the latter PES, quantum classical trajectory (QCT) calculations
were performed to determine CRC for the 45$^{th}$ first energy levels 
of H$_2$O for both ortho-- and para--H$_2$ \citep{faure2007}. The range of temperature
covered by these calculations is 100--2000 K (note that CRC are provided 
below 100 K, making use of the quantum CRC of \cite{dubernet2006} for the 
transitions that involve the lowest
five energy levels of either o--H$_2$O or p--H$_2$O and assuming a constant temperature 
dependance for the other transitions.)
Making use of laboratory measurements for the vibrational 
relaxation of water and QCT calculations \citep{faure2005}, the QCT CRC of \citet{faure2007}, obtained for the vibrational ground state,
have subsequently been scaled to provide ro--vibrational CRC for 
the $5^{th}$ first vibrational states \citep{faure2008}. 
Finally, quantum calculations were performed for the H$_2$O -- H$_2$ system 
making use of the same 5D PES than the one used in the QCT calculations. These 
quantum calculations 
provide CRC for the first 45$^{th}$ energy levels of both 
ortho and para H$_2$O and for temperatures covering the 5--1500K range \citep{dubernet2009,daniel2010,daniel2011}.
In these latter studies, an emphasis is made on the inclusion of the excited 
energy levels of H$_2$. Therefore, apart from the usual state--to--state CRC commonly calculated
in quantum studies (i.e. with H$_2$ remaining in its fundamental rotational level during the collision),
the availability of the information relative to the H$_2$ excitation is used to determine thermalized CRC. 

Owing to the time at which the different CRC sets were made available and owing
to the number of energy levels considered and to the temperature coverage, 
the analysis of water excitation has mainly been based on three sets: the quantum 
H$_2$O -- He CRC of \citet{green1993}, the QCT CRC of \citet{faure2007} and the quantum
CRC of \citet{dubernet2006,dubernet2009,daniel2010,daniel2011}. In what follows,
we discuss the H$_2$O line intensity predictions based on these three CRC sets and
calculated with a precise non--LTE non--local radiative transfer code. 
Additionally, we choose to focus the discussion only on the o--H$_2$O 
symmetry since the results are similar for the p--H$_2$O symmetry.

\subsection{Radiative transfer modelling}

We performed RT calculations using the various sets 
of CRC available for o--H$_2$O and p--H$_2$O.
The numerical code used to solve the molecular excitation and the 
RT problem is described in \citet{daniel2008}.
The water vapor spectroscopic parameters, i.e. line frequencies and Einstein coefficients, are 
taken from the HITRAN database \citep{rothman2009}.
The model consists of a static spherical homogeneous cloud with 
turbulence velocity dispersion fixed at 1 km s$^{-1}$ and with a radius of 
$\sim 4.5 \, 10^{16}$ cm (i.e. 6'' at 500 pc). Since we focus 
on collisional effects, pumping by 
the dust infrared radiation is not included in a first stage, so that
the population of the water energy levels is only due to the collisions 
with the H$_2$ molecules and to radiative trapping due to line opacity effects.
The inclusion of dust emission is however briefly discussed in Sect. \ref{discussion}.
A grid of models has been run leaving the 
gas temperature, the H$_2$ volume density and the water abundance relative to H$_2$ as free parameters.
Those quantities vary in the ranges : 
T $\in$ [200K;1000K], n(H$_2$) $\in$ [10$^6$ ; 2 $\, 10^9$] cm$^{-3}$ and 
$\chi$(H$_2$O) $\in$ $\left\{ 10^{-8} ; 10^{-6} ; 10^{-4} \right\}$, for both o--H$_2$O and p--H$_2$O.
All the RT models are performed considering the first 45$^{th}$ energy levels of
p--H$_2$O or o--H$_2$O, irrespective of the CRC set used.

Since we solve the non--local excitation problem, we consider
an average of the parameters that describe the radiative transitions 
in order to compare the results based on the different CRC sets. 
Therefore, the influence of the CRC on the line emission
is estimated from the quantity : 
\begin{equation}
\bar{I_j} = B(T_{bg}) \, e^{-\tau_j} + B(\bar{T}_{ex}) \times (1 - e^{-\tau_j}) 
\label{eq1}
\end{equation}
where $B(T)$ stands for the Planck function, $\tau_j$ for the opacity at the $j$ line center
and $T_{bg}$ is the temperature of the background radiation (set to the CMB temperature 
in what follows, except if specified).
$\bar{I}_j$ corresponds to the specific intensity
of the transition $j$, along a ray with constant excitation conditions 
(i.e. constant $\bar{T}_{ex}$).
The excitation temperature $\bar{T}_{ex}$ is
defined as an average over the $N$ radial grid points of the models, and is
calculated according to :
\begin{equation}
\bar{T}_{ex} = \frac{\mathcal{A}}{2} \sum_{i=1}^{N-1} \left[ \kappa(r_i) \, T_{ex}(r_{i+1})+\kappa(r_{i+1}) \, T_{ex}(r_{i+1}) \right] \times \frac{r_{i+1}-r_{i}}{r_{N}-r_{1}}
\end{equation}
where $r_i$ stands for the distance of the $i^{th}$ grid point to the centre of the sphere,
$\mathcal{A}$ is a normalisation coefficient and $\kappa(r)$ is the $j$ line 
absorption coefficient at radius $r$. Calculating $\bar{T}_{ex}$ this way, we
prevent that lines with suprathermal excitation (i.e. $T_{ex}$ $>$ $T_K$),  
but with nearly equal populations for the upper and lower levels,
have their averaged $\bar{T}_{ex}$ overestimated.
Indeed, even under homogeneous conditions, most of the lines show 
large variations of $T_{ex}(r)$ at the edge of the sphere, which are in general
associated with low values of the absorption coefficients. Weighting $T_{ex}(r)$
by the associated $\kappa(r)$ enables to reduce the influence of such variations
and insure that its mean value is representative of the volume of 
the cloud which emits photons.
The line opacity $\tau_j$ is given by :
\begin{equation}
\tau_j = \frac{1}{2} \sum_{i=1}^{N-1} \left[ \kappa(r_i)+\kappa(r_{i+1}) \right] \times \left( r_{i+1}-r_{i} \right)
\end{equation}

The comparison of the various models is done using $\bar{I}_j$ rather than using the line intensity peaks
(noted $max(I_j$) in what follows)
or the integrated area of the lines (noted $W_j$), for various reasons. At a first glance, the two latter 
estimators would result more natural, since such quantities would be the one used to compare
observations and models. However, a difficulty arise from the fact that for a given line, 
the line profiles may differ from one model to the other. In such a case, the ratio obtained considering
either $max(I_j$) or $W_j$ can differ by a few 10\%. 
 
\begin{figure}[h]
\begin{center}
\includegraphics[angle=0,scale=0.45]{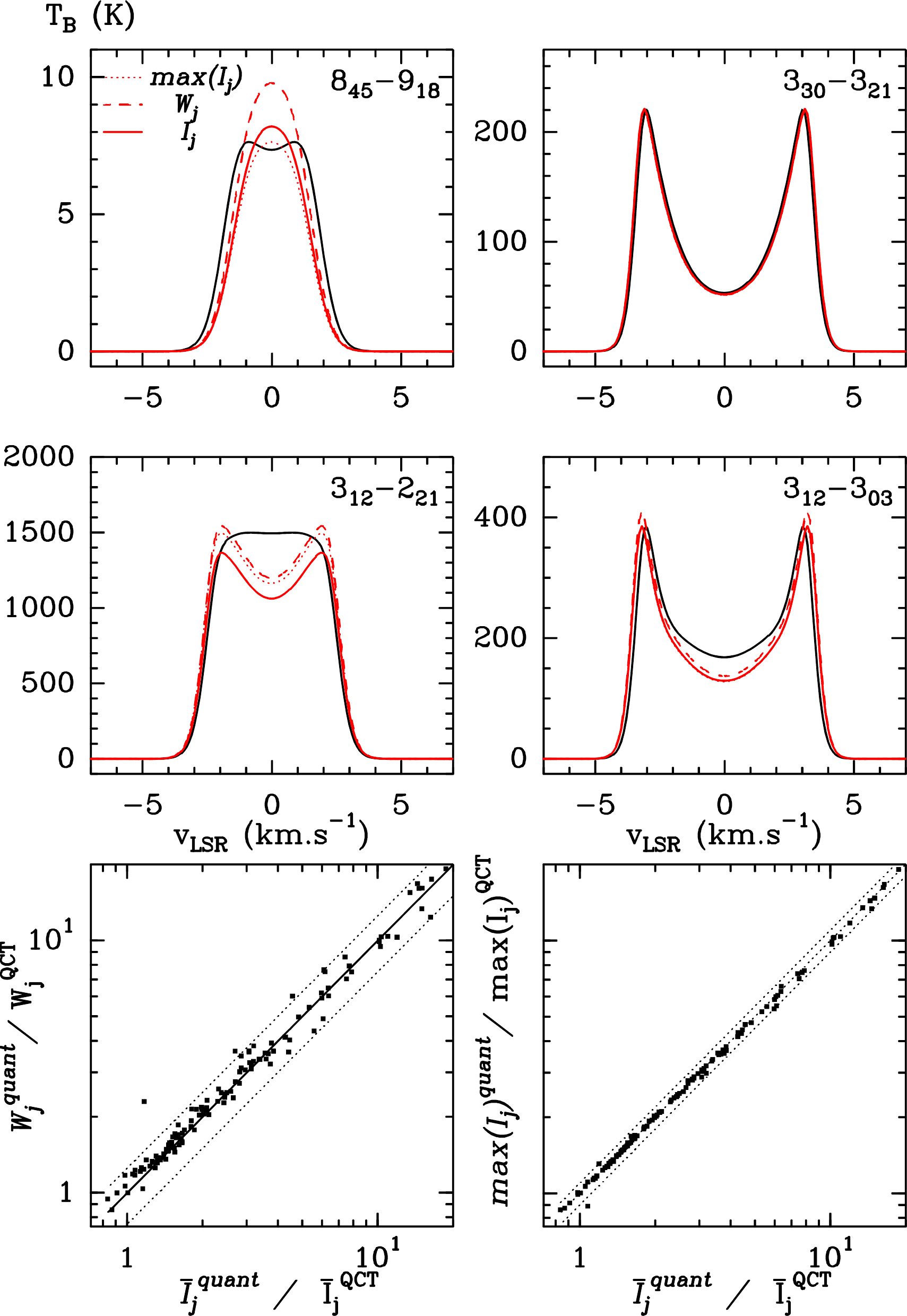}
\end{center}
\caption{%Comparison of the line intensity ratios based on three estimators 
%(see text for the details of the models being compared).  For a given line $j$, the ratio is 
%determined using either $\bar{I}_j$, the integrated area of the line noted $W_j$,
%or the peak intensity of the line noted $max(I_j)$.
%
Comparison of the line intensity ratios based on three estimators: 
$\bar{I}_j$, the integrated area $W_j$,
or the peak intensity $max(I_j)$ (see text for details) .
The four upper panels give the brightness temperature of a few o--H$_2$O lines. The black curve
corresponds to the profile obtained with the quantum CRC set and the red curves to the QCT CRC 
set (see text for details). In the latter case, the profile is scaled according to the ratios of
$\bar{I}_j$ (plain curve), $W_j$ (dashed curve) or $max(I_j)$ (dotted curve) so that the profiles which
are plotted would correspond to a ratio of unity according to the three estimators.
The two bottom panels represent the variations of the ratios obtained between the various indicators 
and for all the lines of the model.
}
\label{comp100}
\end{figure}

This is illustrated in Fig. \ref{comp100}. In this figure,
the models considered correspond to the physical parameters : T$_K$ = 800 K, $n$(H$_2$) = 2 10$^7$ cm$^{-3}$
and $\chi$(o-H$_2$O) = 10$^{-6}$. The collisional partner is p-H$_2$ and the results compared are 
obtained using the QCT CRC of \citet{faure2007} 
and STS quantum CRC of \citet{dubernet2009} (respectively labelled \textit{QCT} and \textit{quant.} in the following).
These CRC sets are more extensively discussed in Sect. \ref{section:sets}.
The two bottom panels show the variation of $W_j^{quant.}/W_j^{QCT}$ (left panel) and 
$max(I_j)^{quant.}/max(I_j)^{QCT}$ (right panel) as a function of $\bar{I_j}^{quant.}/\bar{I_j}^{QCT}$.
Considering the integrated intensity, it can be seen that most of the $W_j$ and $\bar{I}_j$ ratios show 
maximum differences of the order of 25 \% (the two dashed lines on the left panel correspond to straight
lines of slopes 0.75 and 1.25 and thus delimit a maximum deviation of 25 \%.)
Considering the peak intensity (right panel), it can be seen that the correlation between the 
$max(I_j)$ and $\bar{I}_j$ ratios is better than for the case of the integrated intensity. In that case,
the maximum differences are lower than 10 \% (the dashed lines correspond to straight
lines of slopes 0.9 and 1.1).
The four upper panels illustrate the fact that the three estimators start to lead to different 
ratios when the line profiles of the two models differ. In this figure, we report the brightness 
temperature for the  impact parameter that crosses the center of the sphere. The profile
obtained with the p--H$_2$ QCT CRC correspond to the black curves. The red curves
correspond to the profiles obtained with the p--H$_2$ quantum CRC, and scaled by the ratio
$\bar{I}_j^{QCT}/\bar{I}_j^{quant.}$ (plain curve), $W_j^{QCT}/W_j^{quant.}$ (dashed curve)
and $max(I_j)^{QCT}/max(I_j)^{quant.}$ (dotted curve). In other words, the red curves
would correspond to line profiles that would lead to a ratio of unity, when compared to the black
curve and depending on the estimator used.
It can be seen that for the 
3$_{30}$-3$_{21}$ line, since the line profiles derived from the two CRC sets are identical, the ratio
obtained from the three estimators are identical too. On the other hand, for example for the 
3$_{12}$-2$_{21}$ line, the three estimators lead to different ratios since the line profiles
obtained using the two CRC sets differ.

Finally, we emphasize on the fact that in principle, whatever of the three estimators could be used
in order to perform the comparisons which are presented in the next sections. The conclusions
would not be affected by this choice due to the good correlation between the ratios obtained with those 
three estimators.
We prefer to use $\bar{I}_j$ because of its simplicity and beacuse it reduces the line intensity simply 
to two parameters, the averaged excitation temperature and the line opacity, the first quantity
being a useful indicator on how far the line is from thermalization.

\subsection{Comparison of rate coefficient sets} \label{comparaison_CRC}

In order to compare two CRC sets, noted SET1 and SET2, we consider the values taken 
by the ratios $x_j = \bar{I}^{\, SET1}_j / \bar{I}^{\, SET2}_j$, where $\bar{I}$ is defined by eq. \ref{eq1} 
and where the index $j$ stands for the $j^{th}$ radiative transition. 
The comparisons are made considering the statistics on the 
$M$ radiative lines that respect the criteria:
\begin{itemize}
\item the line does not show substantial population inversion (i.e. we adopt as a selection
criterium that the opacity of a line in the inverted region can not be greater 
than 0.5\% of the opacity of the line in the thermal region.)
\item the upper level of the line has an energy below the N$^{th}$ level (labelled as N$_{max}$ in what follows)
\item the intensity of the line as given by eq. \ref{eq1} is above 10 mK
\end{itemize} 
The first criterium means that the masers are discarded from the statistical analysis. Hence,
the following conclusions do not concern the lines observable with the ALMA 
interferometer. We refer to \citet{Daniel12}
for a discussion concerning the impact of the CRC on these masing lines.
From these $M$ lines, we define
the mean (noted $m$) and standard deviation (noted $\sigma$)
associated of the $x_j$ ratios given by:
\begin{equation}
\sigma = \sqrt{\frac{1}{M} \, \sum_{j=1}^M (x_j - m)^2}
\label{eq100}
\end{equation}
Additionally, we discuss the results using the normalised standard deviation
defined as $\sigma/m$ rather than $\sigma$.

\section{Comparison between various H$_2$O CRC} \label{section:sets}

In what follows, we discuss the H$_2$O line intensity predictions based on the three CRC sets which 
have been widely used, i.e. the quantum H$_2$O -- He CRC of \citet{green1993}, 
the QCT CRC of \citet{faure2007} and the quantum CRC of \citet{dubernet2006,dubernet2009,daniel2010,daniel2011}.

\subsection{o--H$_2$O QCT and quantum STS CRC with H$_2$ compared to quantum He CRC}

In this section, various comparisons are made between the quantum state--to--state (STS) 
CRC calculated with either p--H$_2$ \citep{dubernet2009,daniel2011} or o--H$_2$ \citep{daniel2011}, the quantum STS CRC calculated with He \citep{green1993} and the QCT
calculations \citep{faure2007}. When referring to the quantum H$_2$ STS CRC, it is assumed that the CRC stands 
for the CRC where H$_2$ remains in its fundamental rotational energy level (i.e. either $j_2 = 0$ for p--H$_2$ or $j_2 = 1$ 
for o--H$_2$ with $j_2$ being the H$_2$ rotational quantum number). 
Note that QCT rate coefficients are not STS but obtained for thermal populations of 
p--H$_2$ and o--H$_2$, i.e. they are thermalized CRC (see below).
In the RT calculations,
the CRC calculated with He are scaled according to the 
differing reduced masses of the H$_2$O--He and H$_2$O--H$_2$ systems, 
in order to emulate collisions with H$_2$.

\subsubsection{p--H$_2$ rate coefficients}

\begin{figure*}[h]
\begin{center}
%  \subfigure[]{\includegraphics[angle=270,scale=0.37]{./GRAPHS/XH2O_1-PARA-Daniel_over_He_JMAX=20.pdf}} \quad
%  \subfigure[]{\includegraphics[angle=270,scale=0.37]{./GRAPHS/XH2O_1-PARA-Faure_over_He_JMAX=20.pdf}} \\
%  \subfigure[]{\includegraphics[angle=270,scale=0.37]{./GRAPHS/XH2O_2-PARA-Daniel_over_He_JMAX=20.pdf}} \quad
%  \subfigure[]{\includegraphics[angle=270,scale=0.37]{./GRAPHS/XH2O_2-PARA-Faure_over_He_JMAX=20.pdf}} \\
%  \subfigure[]{\includegraphics[angle=270,scale=0.37]{./GRAPHS/XH2O_3-PARA-Daniel_over_He_JMAX=20.pdf}} \quad
%  \subfigure[]{\includegraphics[angle=270,scale=0.37]{./GRAPHS/XH2O_3-PARA-Faure_over_He_JMAX=20.pdf}} \\
\includegraphics[angle=0,scale=1.0]{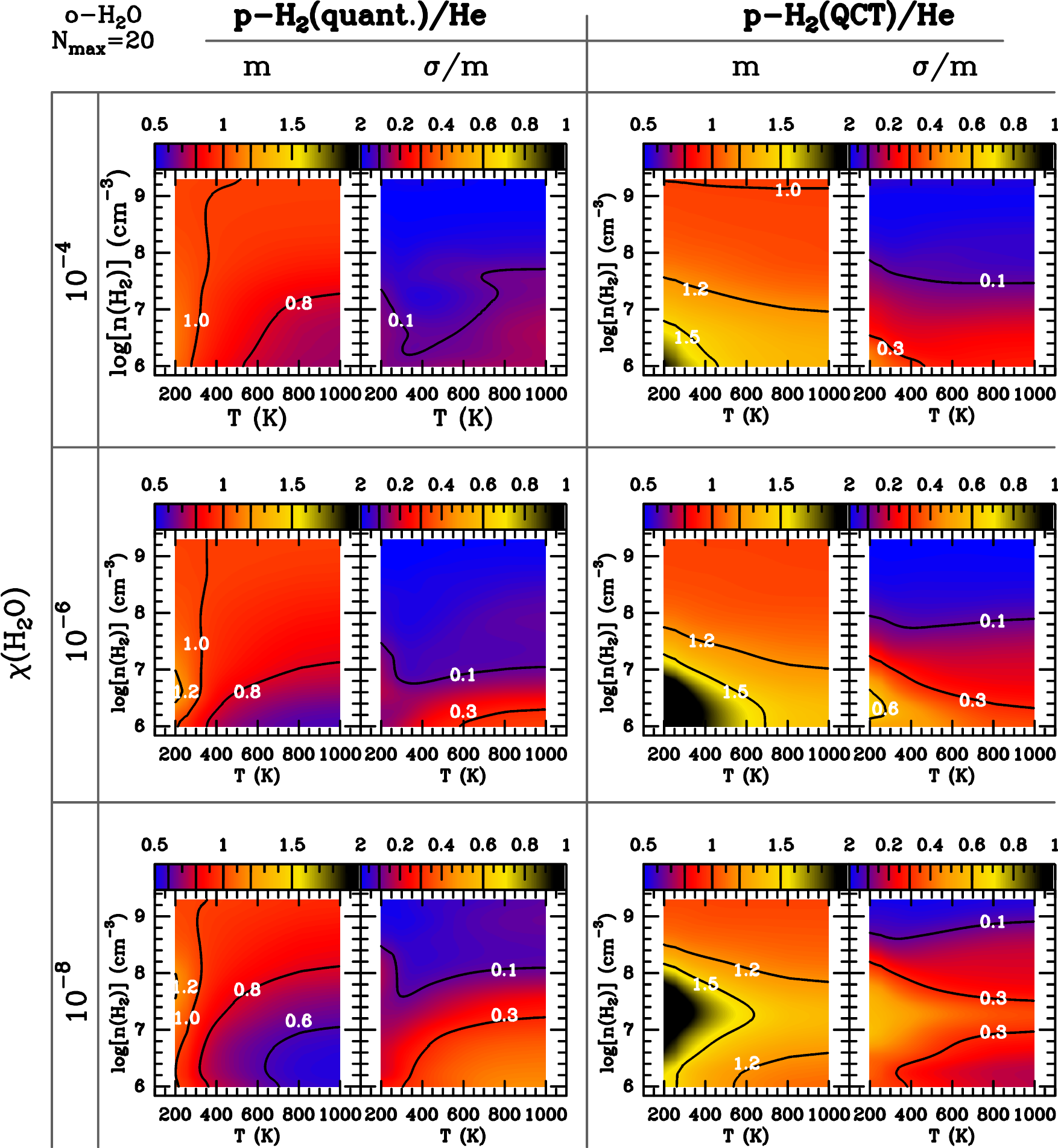}
\end{center}
\caption{Comparison of the mean ($m$) and normalised standard deviations ($\sigma/m$, given by eq. \ref{eq100})
for the ratios $\bar{I}^{quant.}/\bar{I}^{He}$ (left column)
and $\bar{I}^{QCT}/\bar{I}^{He}$ (right column). The intensities $\bar{I}$ are defined by eq. \ref{eq1}. The statistical analysis
is performed on the lines that involve the first N$_{max}$ = 20 o--H$_2$O energy levels.
The comparison deals with p--H$_2$ CRC. 
The quantum CRC with p--H$_2$ are from \citet{dubernet2009,daniel2011}, the quantum CRC with He
from \citet{green1993} and the QCT CRC are from \citet{faure2007}.
In the case of 
the quantum CRC, the rates considered are the STS CRC with H$_2$ in $j_2 = 0$. The comparison is performed at
the water abundances $\chi$(H$_2$O) = $10^{-8}$, $10^{-6}$ and $10^{-4}$} (rows).
\label{comp1}
\end{figure*}

A first comparison is made between the o--H$_2$O / p--H$_2$ CRC sets obtained
with the quantum calculations  and the QCT calculations, both with respect
to the quantum calculations performed with He. 
The mean and normalised standard deviations of the ratios  $\bar{I}^{H_2}/\bar{I}^{He}$ are represented on Fig. \ref{comp1}
for various water abundances and considering the o--H$_2$O lines with upper energy levels below 
the 20$^{th}$ level ($E_u$ $<$ 900 K). 
It appears that the behaviour of the $x_j$ ratios can be distinguished
between two regimes, which are separated by a threshold (noted $n$ in what follows)
in the water volume density $n$(H$_2$O) = $n$(H$_2$) $\times$ $\chi$(H$_2$O).
For water volume densities $n$(H$_2$) $\times$ $\chi$(H$_2$O) $> n $, the mean value is around 1 and the normalised standard deviation
takes low values (typically below 0.1). This corresponds to the \textit{thermalized regime}
where the line intensity ratios are basically independent 
of the adopted CRC set. In this regime, most of the lines show 
large optical depths. These large optical depths imply 
that the critical densities of the lines become
lower than in the optically thin limit\footnote{i.e. $n_c = A_{ul} \left< \beta \right> / C_{ul}$  where $\beta$ is the probability that a photon 
escapes the medium ; with $\beta = 1$ in the optically thin case and $\beta \sim 1/\tau$ when the line becomes
optically thick}, hence producing the thermalization of the level populations.
For $n$(H$_2$) $\chi$(H$_2$O) $< n $, the level populations are determined by both collisional and radiative
processes, making the mean value differing from 1 and leading to an increase of the normalised standard deviation 
(in what follows, this regime is referred as \textit{subthermal regime}).

In Fig. \ref{comp1} we show a comparison of the line intensity ratios 
obtained with the quantum p--H$_2$ CRC (left column) and
QCT p--H$_2$ CRC (right column), both compared to the quantum He CRC.
Examining the p--H$_2$ CRC, we see that 
independently of the abundance considered, the mean value 
is in the range $0.45 < m < 1.35$. The normalised standard deviation is below 0.5 for all the 
free parameter space, which means that most of the lines (roughly 70\% of the lines considered)
show deviations of less than 50\% around the mean value. 
In other words, the main effect introduced by considering the H$_2$ 
CRC will be to scale the line intensities with respect to the intensities derived 
from the He CRC. The maximum difference in the scaling factor is encountered 
at high temperature and low H$_2$ volume densities, where the line intensities based
on the H$_2$  CRC are found to be lower by a factor around $\sim$ 2, for $\chi$(H$_2$O) = 10$^{-8}$.
Additionally, the relative line intensities are expected to vary 
from one set to the other, with maximum variations of the order of 50\% for the bulk of the lines.

The QCT calculations show slightly larger differences, as expected since they 
correspond to thermalized CRC. Indeed, 
we obtain that irrespective of the water vapour abundance,
the mean value of the ratios is in the range $ 1 < m < 2.9$,  for the  
whole parameter space considered here. The normalised
standard deviation can take values of up to $\sim 0.7$.
The spread of the ratio around the mean value is of the same order than the one found 
for the quantum p-H$_2$ CRC.

In Fig. \ref{comp3} (right column), we show a direct comparison of the results predicted by the STS and QCT CRC.
Independently of the water abundance, the QCT CRC predict larger intensities 
(a factor 2 higher) than the STS CRC. This is due to the fact that the QCT calculations correspond
to a thermal average over the H$_2$ energy levels and because of the large differences of the CRC for H$_2$ in its 
fundamental state $j_2 = 0$ and in the $j_2 = 2$ state. This is further discussed when considering the thermalized 
CRC (see Sect. \ref{thermalized}).

\subsubsection{o--H$_2$ rate coefficients}

\begin{figure*}[h]
\begin{center}
  % \subfigure[]{\includegraphics[angle=270,scale=0.37]{./GRAPHS/XH2O_1-ORTHO-Daniel_over_He_JMAX=20.pdf}} \quad
  % \subfigure[]{\includegraphics[angle=270,scale=0.37]{./GRAPHS/XH2O_1-ORTHO-Faure_over_He_JMAX=20.pdf}} \\
  % \subfigure[]{\includegraphics[angle=270,scale=0.37]{./GRAPHS/XH2O_2-ORTHO-Daniel_over_He_JMAX=20.pdf}} \quad
  % \subfigure[]{\includegraphics[angle=270,scale=0.37]{./GRAPHS/XH2O_2-ORTHO-Faure_over_He_JMAX=20.pdf}} \\
  % \subfigure[]{\includegraphics[angle=270,scale=0.37]{./GRAPHS/XH2O_3-ORTHO-Daniel_over_He_JMAX=20.pdf}} \quad
  % \subfigure[]{\includegraphics[angle=270,scale=0.37]{./GRAPHS/XH2O_3-ORTHO-Faure_over_He_JMAX=20.pdf}} \\
  \includegraphics[angle=0,scale=1.0]{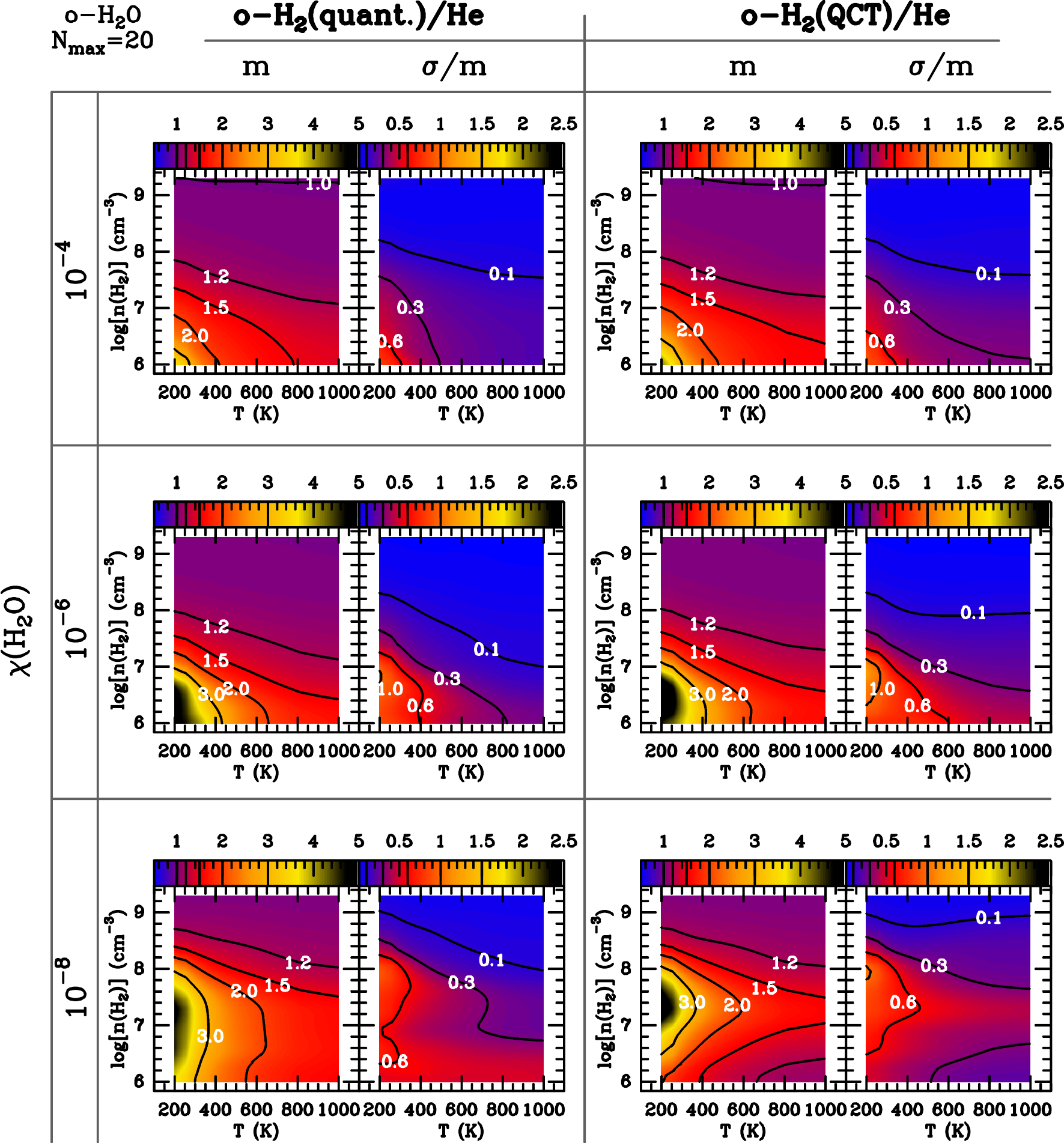}
\end{center}
\caption{Same as Fig. \ref{comp1} but for the collisions that involve o--H$_2$.}
\label{comp2}
\end{figure*}

A second comparison concerns the results obtained with o--H$_2$ as a collisional partner.
The results obtained with the quantum and QCT CRC are compared to the results obtained
with He in Fig. \ref{comp2}. Qualitatively, the H$_2$ quantum and QCT calculations 
compares similarly to the results obtained with He.
The main effect concerns the overall scaling of the ratios.  In the subthermal regime, the intensity ratios obtained using o--H$_2$ 
are higher than the one obtained using He, irrespective of the water abundance. A typical increase of 50\% is 
found for the intensity of the lines, with mean values that can be higher than 3 in the regime 
of low temperatures and low densities. This result is expected from a direct consideration of CRC
obtained with o--H$_2$. Indeed, these rates are typically higher than the rates with He, 
by factors of up to a factor 10. So, the water energy levels are more easily populated when considering o--H$_2$ 
as a collisional partner. This results in brighter lines.
Additionally, we note that the normalised standard deviation is high and its variations are
correlated with the variations of the mean value, i.e. the higher the mean value is, 
the higher the normalised standard deviation is. 

\begin{figure*}[ht]
\begin{center}
%  \subfigure[]{\includegraphics[angle=270,scale=0.37]{./GRAPHS/XH2O_1-ORTHO-D_over_F_JMAX=20.pdf}} \quad
  %\subfigure[]{\includegraphics[angle=270,scale=0.37]{./GRAPHS/XH2O_1-PARA-D_over_F_JMAX=20.pdf}} \\
  %\subfigure[]{\includegraphics[angle=270,scale=0.37]{./GRAPHS/XH2O_2-ORTHO-D_over_F_JMAX=20.pdf}} \quad
  %\subfigure[]{\includegraphics[angle=270,scale=0.37]{./GRAPHS/XH2O_2-PARA-D_over_F_JMAX=20.pdf}} \\
  %\subfigure[]{\includegraphics[angle=270,scale=0.37]{./GRAPHS/XH2O_3-ORTHO-D_over_F_JMAX=20.pdf}} \quad
  %\subfigure[]{\includegraphics[angle=270,scale=0.37]{./GRAPHS/XH2O_3-PARA-D_over_F_JMAX=20.pdf}} \\
  \includegraphics[angle=0,scale=1.0]{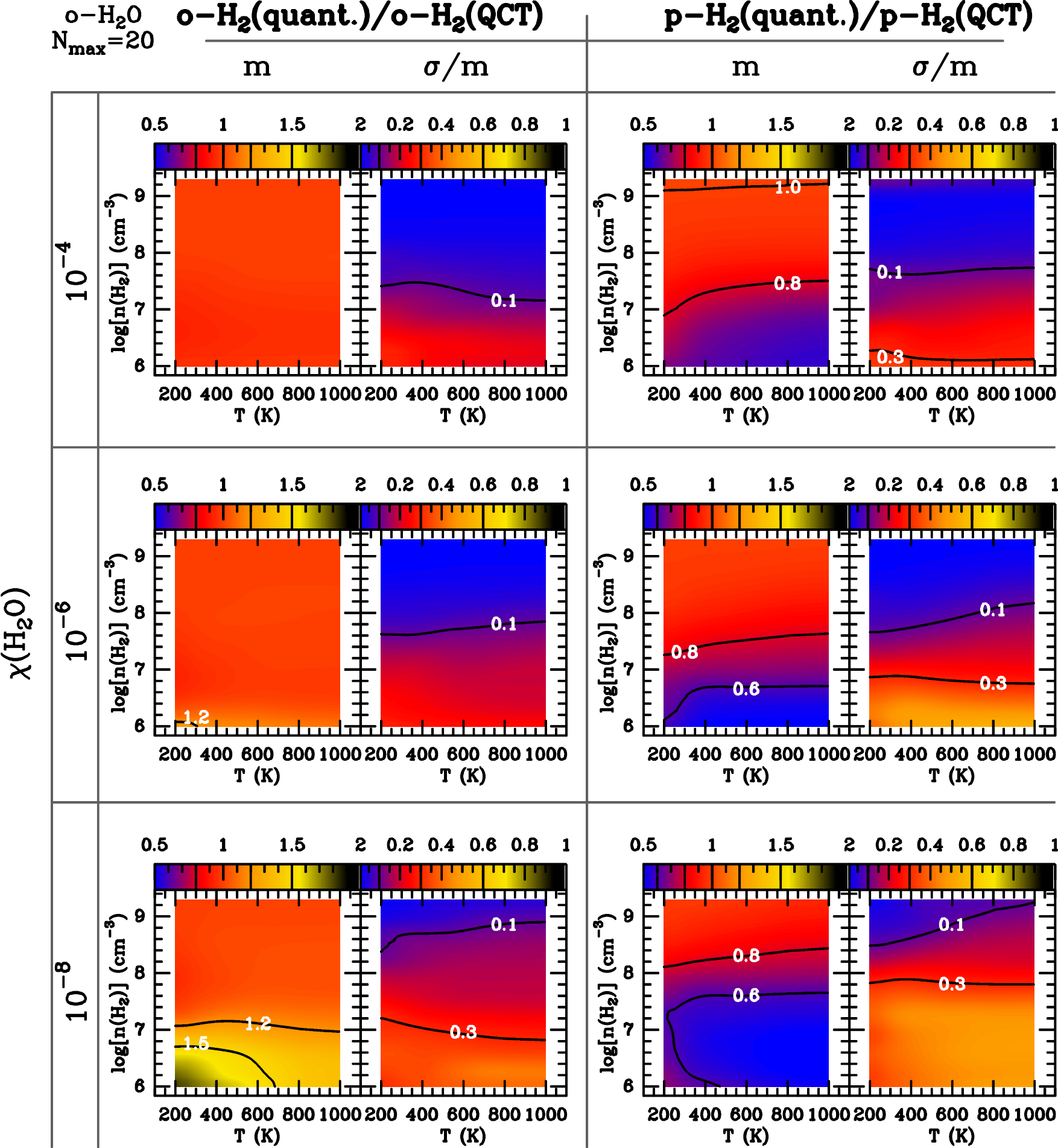}
\end{center}
\caption{Comparison of the results based on the quantum CRC from \citet{dubernet2009,daniel2011}
with the QCT CRC from \citet{faure2007}, for both o--H$_2$ (left column) and p--H2 (right column).}
\label{comp3}
\end{figure*}

A direct comparison of the results obtained with the quantum and QCT calculations is shown in Fig. \ref{comp3}
for o--H$_2$ (left column). The differences found for the intensities
are modest as long as the water abundance is such
that $\chi$(H$_2$O) $\geq 10^{-6}$. In that case, the mean value is found to be around 1 
and the normalised standard deviation is below 0.3 for all the parameter space.
The main differences are found  for n(H$_2$) $< 10^7$ cm$^{-3}$
and $\chi$(H$_2$O) $\sim 10^{-8}$ cm$^{-3}$ where mean values 
in the range $1.5 < m < 2$ are obtained.

\subsection{Quantum thermalized CRC with QCT or STS CRC} \label{thermalized}

In this section, we compare the intensities predicted using the 
quantum state--to--state CRC with the one derived using the thermalized CRC \citep[both defined in][]{dubernet2009,daniel2011}, defined as :
\begin{eqnarray}
R_{ij} = \sum_{j_2} n(j_2) \sum_{j_2'} C_{ij}(j_2 \to j_2')
\end{eqnarray}
In this expression, $C_{ij}(j_2 \to j_2')$ stands for the state--to--state CRC
from state $i$ to state $j$ for the H$_2$O molecule and that corresponds
to the transition $j_2 \to j_2'$ for the H$_2$ molecule.
These thermalized CRC thus consider the possibility of energy transfer for both the target 
molecule (H$_2$O in the present case) and the collider.
The populations of the H$_2$ energy levels [noted $n(j_2)$ in the above expression] are assumed to be in thermal equilibrium, so that
the populations are given by the Boltzmann distribution. In principle, any astrophysical study should consider thermalized
rather than STS CRC when dealing with line excitation. In practice, the H$_2$O molecule 
\citep[and its isotopomers in][]{faure2012} is the only molecule 
for which thermalized CRC have been calculated with a quantum approach. In what follows, we will 
emphasise on the effects introduced by considering thermalized rather than STS CRC, keeping in mind 
that the current findings obtained for the case of water vapour can be extrapolated to other molecules. \newline

\begin{figure}[h!]
\begin{center}
  \includegraphics[angle=270,scale=0.35]{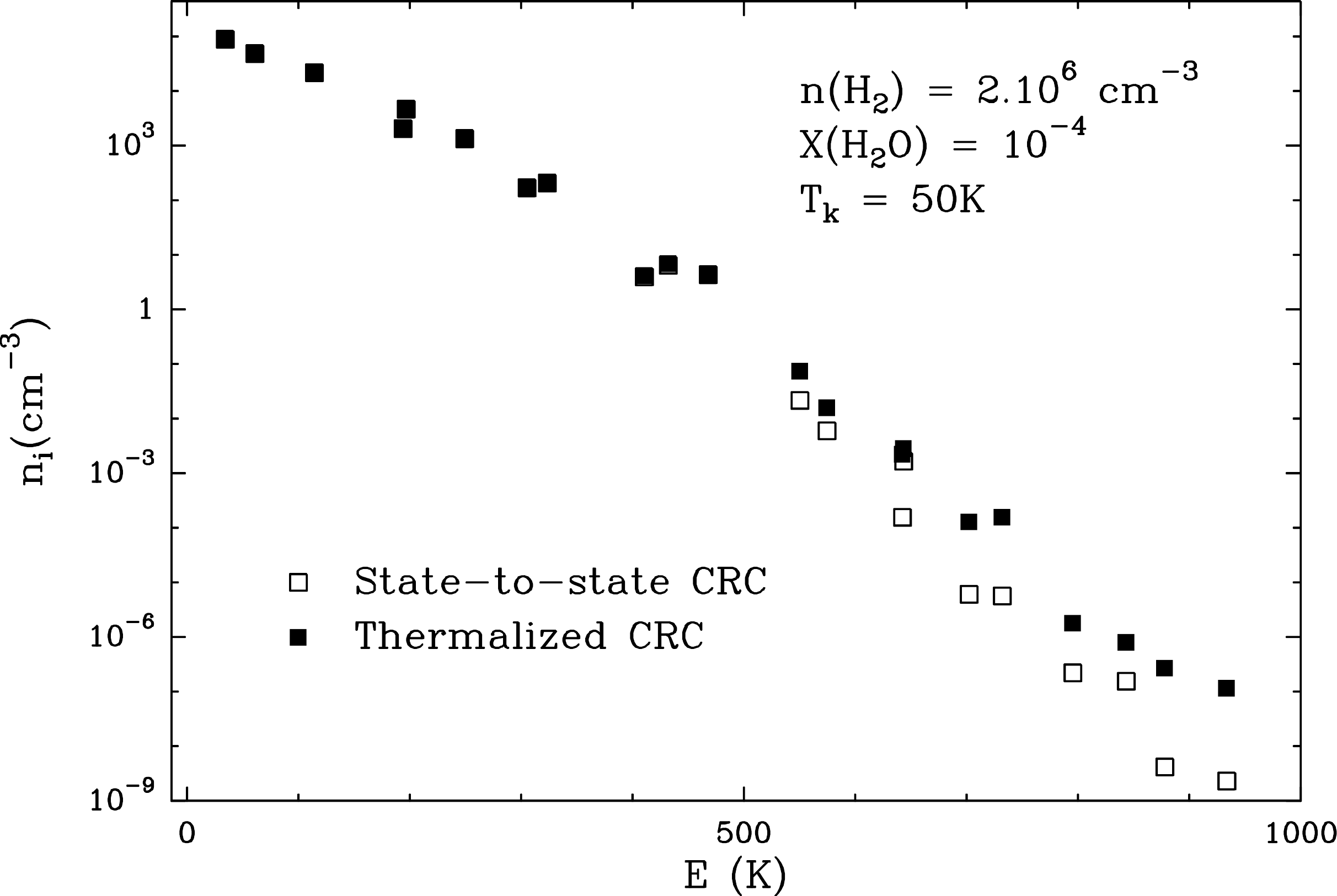}
  \end{center}
\caption{Populations of the o--H$_2$O energy levels as a function of the level energy, for a model
with parameters : n(H$_2$) = $10^6$ cm$^{-3}$, T$_K$ = 50 K and $\chi$(H$_2$O) = $10^{-4}$. The populations
obtained with the state--to--state CRC correspond to the open boxes whereas the one obtained with the thermalized
CRC correspond to the filled boxes.}
\label{pop}
\end{figure}

The thermalized CRC
differ from the quantum STS CRC in two ways.
In the following discussion, we consider the case of the collisions with p--H$_2$.
At low temperature (i.e. T $\leq$ 50 K), only
the fundamental level of the p--H$_2$ molecule is substantially populated. The thermalized CRC
thus reduce to $R_{ij} \sim C_{ij}(0 \to 0) + C_{ij}(0 \to 2)$. At low temperature, the second term of this expression
is often negligible compared to the first term, except for the H$_2$O transitions 
for which the variation of the energy induced by the collision is higher than the energy necessary 
to excite the p--H$_2$ molecule to its first excited state. In the case of p--H$_2$, this corresponds to transitions
that satisfy $\Delta E_{ij} >$ 500 K. For these transitions, the term $C_{ij}(0 \to 2)$ starts to be 
dominant in the evaluation of the thermalized CRC since such a rate can be higher 
by up to a factor 10 in comparison to the STS 
CRC where H$_2$ remains in its fundamental level. The main effect induced by this process
is to efficiently populate the H$_2$O energy levels for the levels with energies 
higher than 500 K.  This is illustrated in Fig. \ref{pop}, for a model with parameters 
$n$(H$_2$) = $10^6$ cm$^{-3}$, T$_K$ = 50 K and $\chi$(H$_2$O) = $10^{-4}$. 
In this figure, we report the water vapour level populations averaged over radius.
It can be seen 
that for the levels with energies below 500 K, the STS and thermalized CRC give similar results for the 
level populations. On the other hand, for the levels with energies higher than 500 K, the populations obtained using
the thermalized CRC are globally 10 times higher than the one obtained from the STS CRC.
Note that in the case of low temperatures, these differences are irrelevant for any astrophysical study since the levels 
higher than 500 K correspond to transitions which are far below the detection limit of the current telescopes.
This example is however given to emphasize the effect introduced by the terms $C_{ij}(0 \to 2)$, 
the same effect being found at higher temperatures. However, at higher temperatures, the transitions 
from the $j_2=2$ state make the influence of those terms less evident when 
considering the level populations, as discussed below.

At higher temperature, the first p--H$_2$ excited state starts to be substantially populated.
Collisions from the state $j_2 = 2$ thus influence the evaluation of the thermalized CRC.
In this case, all the H$_2$O transitions are affected by the scaling of the CRC
due to the term $n(j_2=2) \times C_{ij}(2 \to 2)$. In the case of the H$_2$O molecule, the term
$C_{ij}(2 \to 2)$ is globally larger than $C_{ij}(0 \to 0)$ by a factor that ranges from 2 to 10 depending 
on the transition. Consequently, the thermalized CRC
for all the transitions are globally increased since at the temperatures 
of $100, 200, 500$ K, the population of the $j_2=2$ state account for 
3, 28 and 58\% of the total p--H$_2$ molecules respectively. As an example, at 200 K, all the thermalized CRC
are increased by factors in the range 1.5--3.5 depending on the transition.
Additionally, as discussed in the case of the low temperatures, the CRC for the transitions 
with $\Delta_{ij} > 500 K$ will be increased due to the term $C_{ij}(0 \to 2)$.  
 
\begin{figure}[h!]
\begin{center}
 \includegraphics[angle=270,scale=0.37]{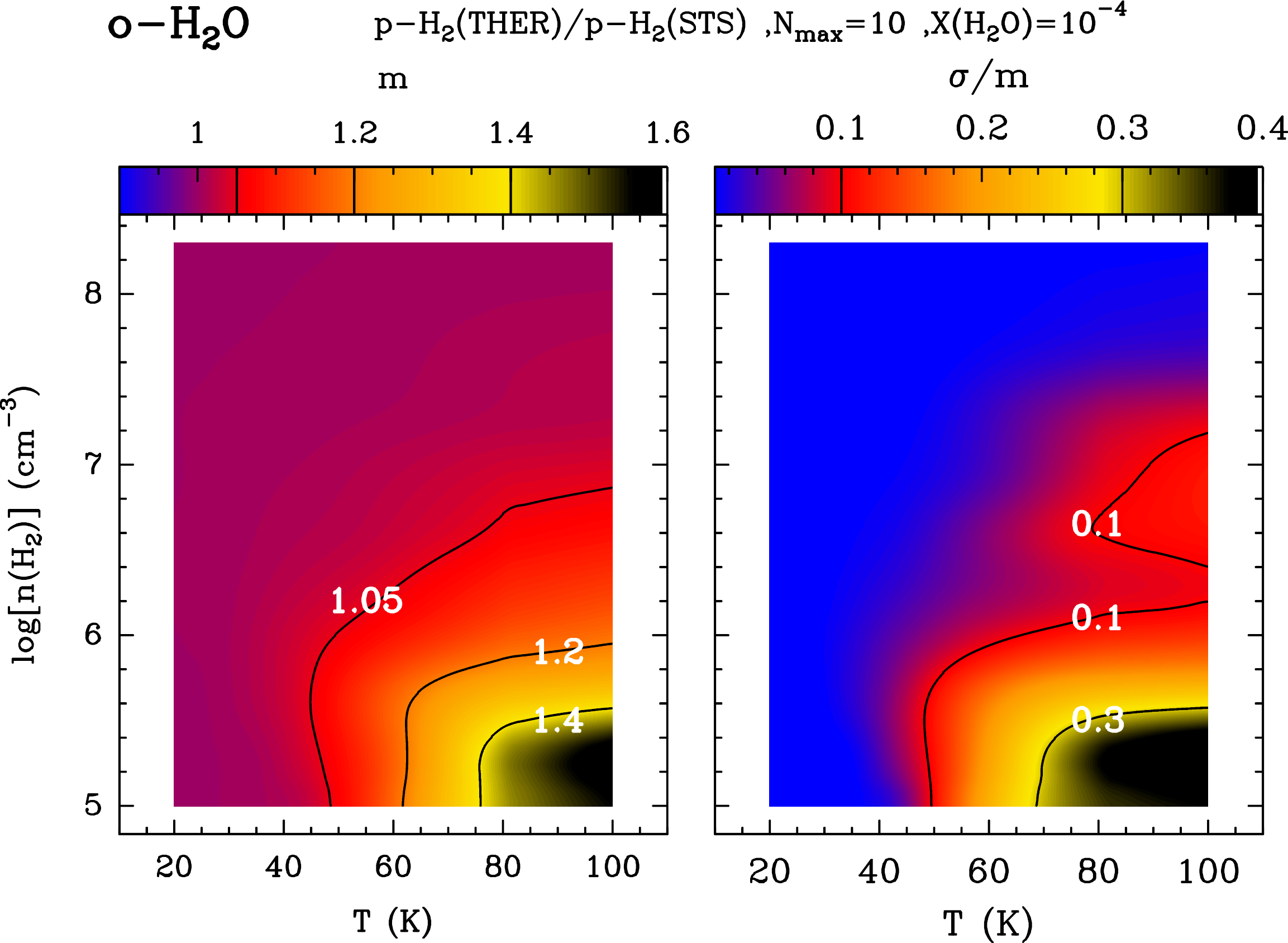} \\
\end{center}
\caption{Comparison of the results based on the thermalized and STS CRC from \citet{dubernet2009,daniel2011} for gas temperatures in the range 20--100 K and
for a water abundance $\chi$(H$_2$O) = 10$^{-4}$.}
\label{comp20}
\end{figure}

Finally, Fig. \ref{comp20} shows a comparison of the results based on thermalized and STS CRC,
for temperatures in the range 20--100 K
and for the 10$^{th}$ first o--H$_2$O rotational energy levels. From this figure, it appears that the temperature 
at which the thermalized CRC start to influence the line intensities is around $T_K \sim 60$ K.
 
\begin{figure*}[h!]
\begin{center}
  %\subfigure[]{\includegraphics[angle=270,scale=0.35]{./GRAPHS/XH2O_1-PARA-THER_over_QCT_JMAX=20.pdf}} \quad
  %\subfigure[]{\includegraphics[angle=270,scale=0.35]{./GRAPHS/XH2O_1-PARA-THER_over_STS_JMAX=20.pdf}} \\
  %\subfigure[]{\includegraphics[angle=270,scale=0.35]{./GRAPHS/XH2O_2-PARA-THER_over_QCT_JMAX=20.pdf}} \quad
  %\subfigure[]{\includegraphics[angle=270,scale=0.35]{./GRAPHS/XH2O_2-PARA-THER_over_STS_JMAX=20.pdf}} \\
  %\subfigure[]{\includegraphics[angle=270,scale=0.35]{./GRAPHS/XH2O_3-PARA-THER_over_QCT_JMAX=20.pdf}} \quad
  %\subfigure[]{\includegraphics[angle=270,scale=0.35]{./GRAPHS/XH2O_3-PARA-THER_over_STS_JMAX=20.pdf}} \\
    \includegraphics[angle=0,scale=1.0]{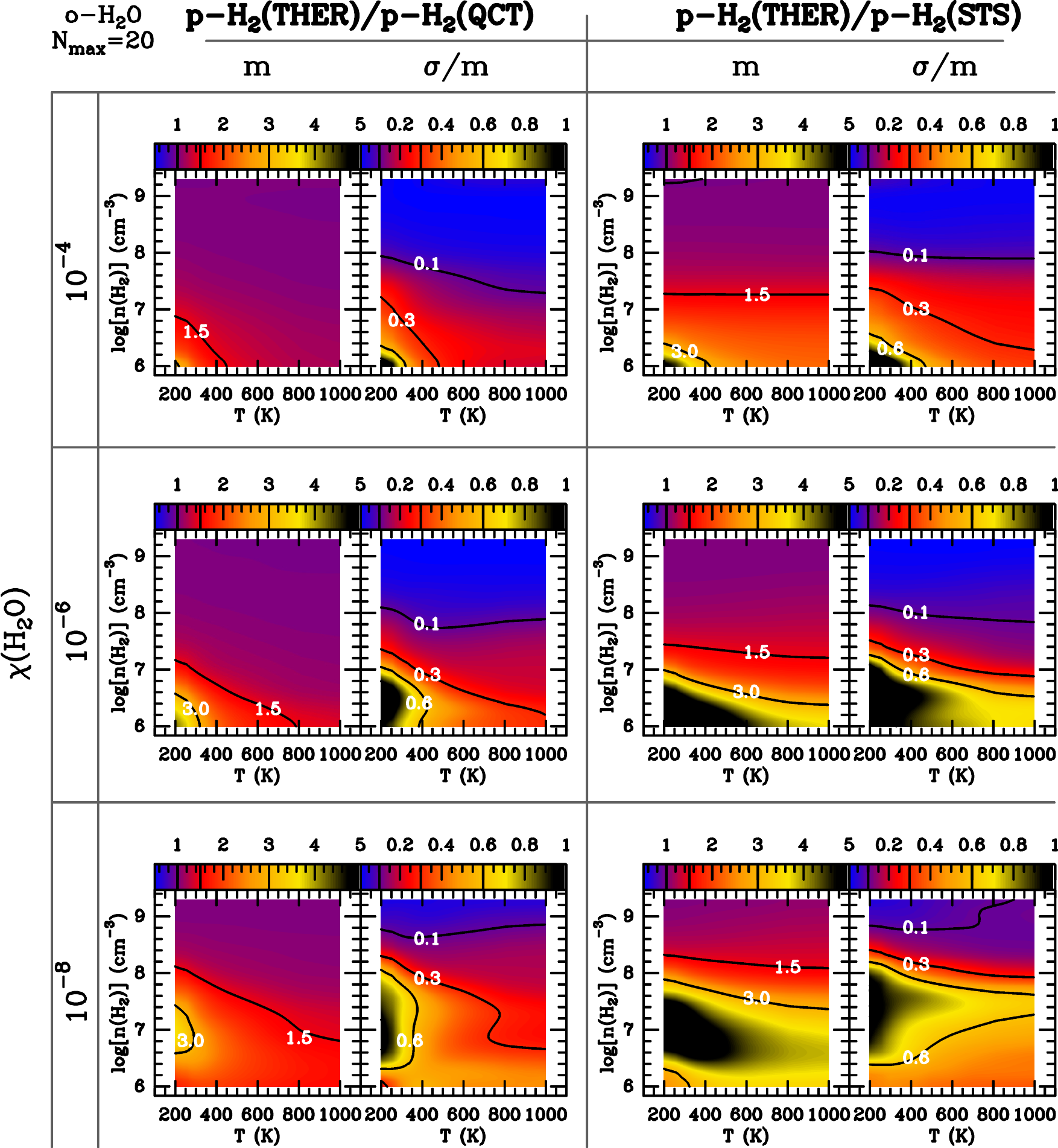}
\end{center}
\caption{Comparison of the results based on the QCT CRC from \citet{faure2007} (left column) 
and state--to--state CRC from \citet{dubernet2009,daniel2011} (right column), 
both with the thermalized CRC from \citet{dubernet2009,daniel2011} for p--H$_2$.}
\label{comp10}
\end{figure*} 

\subsubsection{Comparison of the thermalized, STS and QCT CRC}

In Fig. \ref{comp10}, we compare the quantum STS \citep{dubernet2009,daniel2011}
and QCT \citep{faure2007} CRC with the results obtained with the 
thermalized CRC for p--H$_2$ \citep{dubernet2009,daniel2011}. 
It can be seen that adopting the thermalized CRC strongly influences
the results. Indeed, in the limit of low density (i.e. $n$(H$_2$) $< 10^{7}$ cm$^{-3}$), both STS and QCT
CRC give results that can differ by more than a factor 3, if we consider the mean value. Such high differences
are found over all the temperatures considered here for what concerns the STS CRC. On the other hand, the  differences
between the QCT and thermalized CRC decrease with temperature. Above 400 K, the mean value obtained from the 
comparison of the QCT and thermalized CRC is below 2.

 \begin{figure}[h!]
\begin{center}
 \includegraphics[angle=270,scale=0.37]{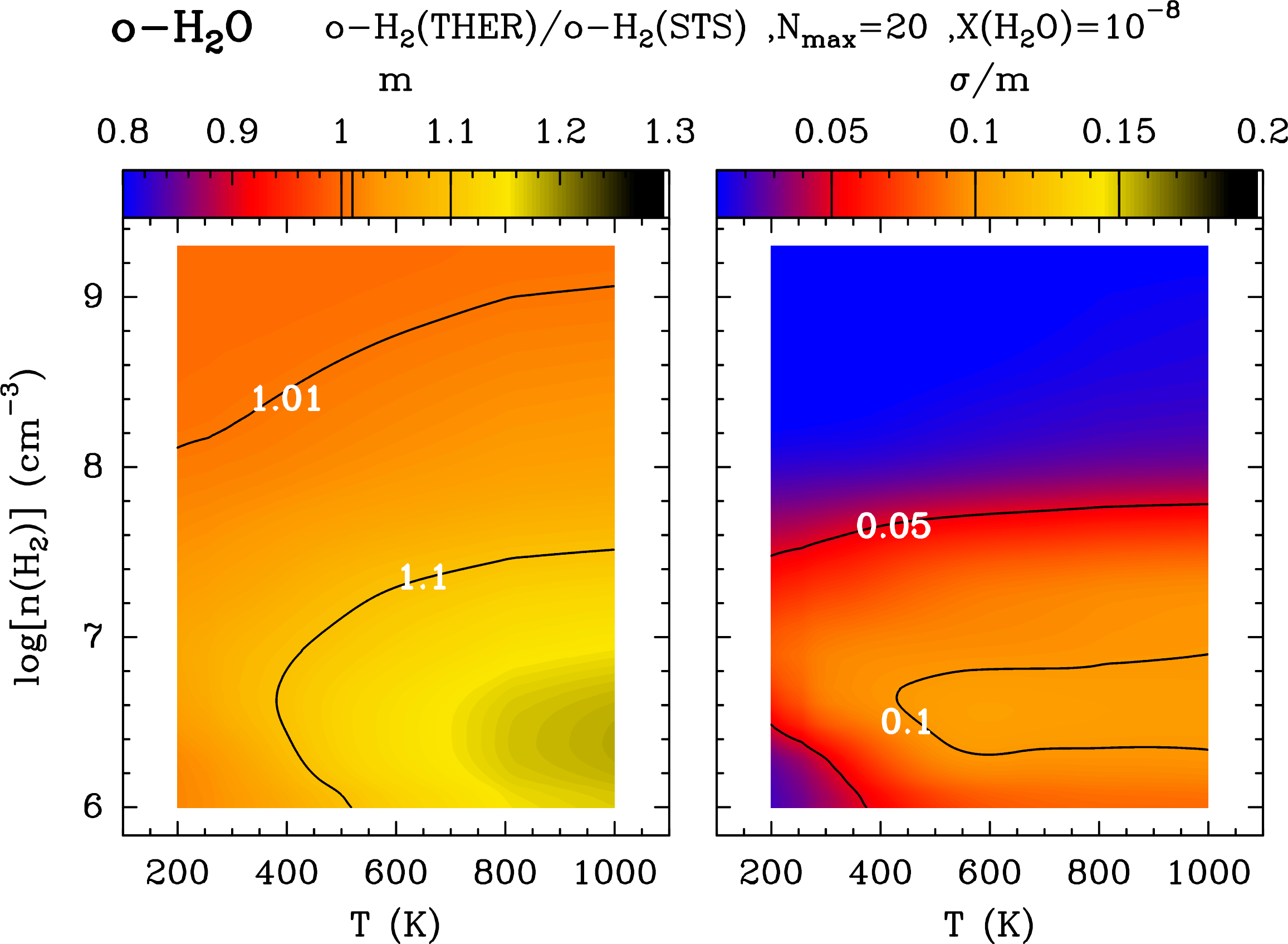} \\
\end{center}
\caption{Comparison of the results based on the thermalized and STS CRC from \citet{daniel2011}, for o--H$_2$ and for gas temperatures in the range 200--1000 K and
for a water abundance $\chi$(H$_2$O) = 10$^{-8}$.}
\label{comp21}
\end{figure} 

In Fig. \ref{comp21}, we compare the results based on thermalized and STS \citep{daniel2011}  
CRC for the o--H$_2$ symmetry.
In this figure, we see that the thermalized CRC give similar results than the STS CRC, with a mean value in the 
range $1 < m < 1.2$ and normalized standard deviation below 0.15 for all the parameter space. The results are shown
for a water abundance $\chi$(H$_2$O) = 10$^{-8}$ but the the results are similar for other abundances. The similarity
of the derived line intensities
is due to the fact that the terms $C_{ij}(j_2 \to j_2')$ that involve the $j_2=1$ state or $j_2 = 3$ state are similar in magnitude.

\begin{figure*}[h!]
\begin{center}
  %\subfigure[]{\includegraphics[angle=270,scale=0.3]{./GRAPHS/H2O_110-101.pdf}} \quad
  %\subfigure[]{\includegraphics[angle=270,scale=0.3]{./GRAPHS/H2O_212-101.pdf}} \\
  %\subfigure[]{\includegraphics[angle=270,scale=0.3]{./GRAPHS/H2O_221-212.pdf}} \quad
   %\subfigure[]{\includegraphics[angle=270,scale=0.3]{./GRAPHS/H2O_303-212.pdf}} \\
   %\subfigure[]{\includegraphics[angle=270,scale=0.3]{./GRAPHS/H2O_312-303.pdf}} \quad
   %\subfigure[]{\includegraphics[angle=270,scale=0.3]{./GRAPHS/H2O_321-212.pdf}} \\
   %\subfigure[]{\includegraphics[angle=270,scale=0.3]{./GRAPHS/H2O_330-221.pdf}} \quad
   %\subfigure[]{\includegraphics[angle=270,scale=0.3]{./GRAPHS/H2O_414-303.pdf}} \\
   \includegraphics[angle=0,scale=1.0]{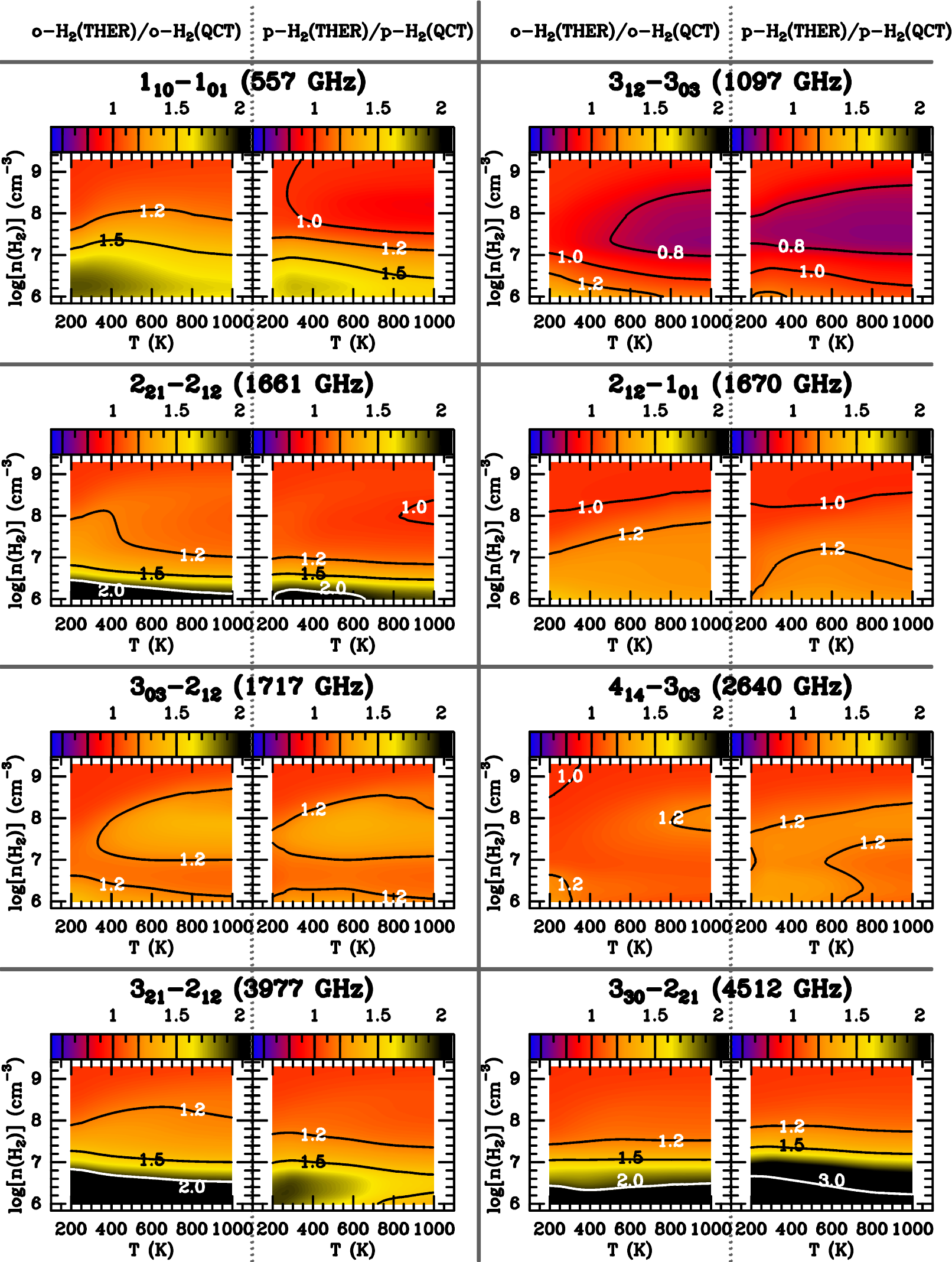}
\end{center}
\caption{Ratio $\bar{I}^{THER}/\bar{I}^{QCT}$ for a few transitions commonly observed
with the Herschel satellite. The thermalized CRC are from \citet{dubernet2009,daniel2011} and
the QCT CRC from \citet{faure2007}.
The ratios are given for collisions that involve o--H$_2$ (left figure)
and p--H$_2$ (right figure) in each panel.}
\label{comp4}
\end{figure*} 

Fig. \ref{comp4} show the $\bar{I}^{THER}/\bar{I}^{QCT}$ ratios for both o--H$_2$ and p--H$_2$,
for eight H$_2$O transitions commonly observed with the HIFI and PACS spectrometers on--board
Herschel . For each line, the ratios for the 
prediction based on the o--H$_2$ set are given in the left panel, while the ratios obtained 
from the CRC that involve p-H$_2$ are given in the right one. 
Note the similarity of the intensity ratio obtained for both the o--H$_2$ and p--H$_2$ symmetries. 
This behaviour will be further discussed in the Sect. \ref{OTPR}.
The figure corresponds to a water abundance $\chi$(H$_2$O) = 10$^{-8}$.
From this figure, it appears that the two CRC sets will lead to qualitatively similar predictions
for the line intensities. Indeed, the maximum variations which are found are of the order of a factor 
of 3, in the worst case. For the lines considered here, and considering collisions with o--H$_2$ 
(the most abundant symmetry for H$_2$ in hot media),
the o--H$_2$O lines that show the largest differences (i.e. greater than a factor of $\sim$ 2) are the 
$1_{10}-1_{01}$, $2_{21}-2_{12}$, $3_{30}-2_{21}$ lines. These differences 
are found at $n$(H$_2$) $< 10^{7}$ cm$^{-3}$ and for all the temperatures considered here.  
On the other hand, some lines show only small variations from 
one set to the other, with intensities that agree within 
20\% (e.g. 2$_{12}$-1$_{01}$ ; 3$_{03}$-2$_{12}$ ; 3$_{12}$-3$_{03}$ ; 4$_{14}$-3$_{03}$).
Moreover, we find that for the majority of the lines, the predictions based 
on the QCT CRC give lower intensities for the o--H$_2$ symmetry 
(for all the lines except the 3$_{12}$-3$_{03}$ for which a ratio between 0.6 and 0.8
are found at high temperatures). This implies that an analysis based on the QCT 
CRC will tend to underestimate the water abundance.

\subsubsection{Behaviour of the thermalized CRC}
\begin{figure}[h!]
\begin{center}
\includegraphics[angle=270,scale=0.35]{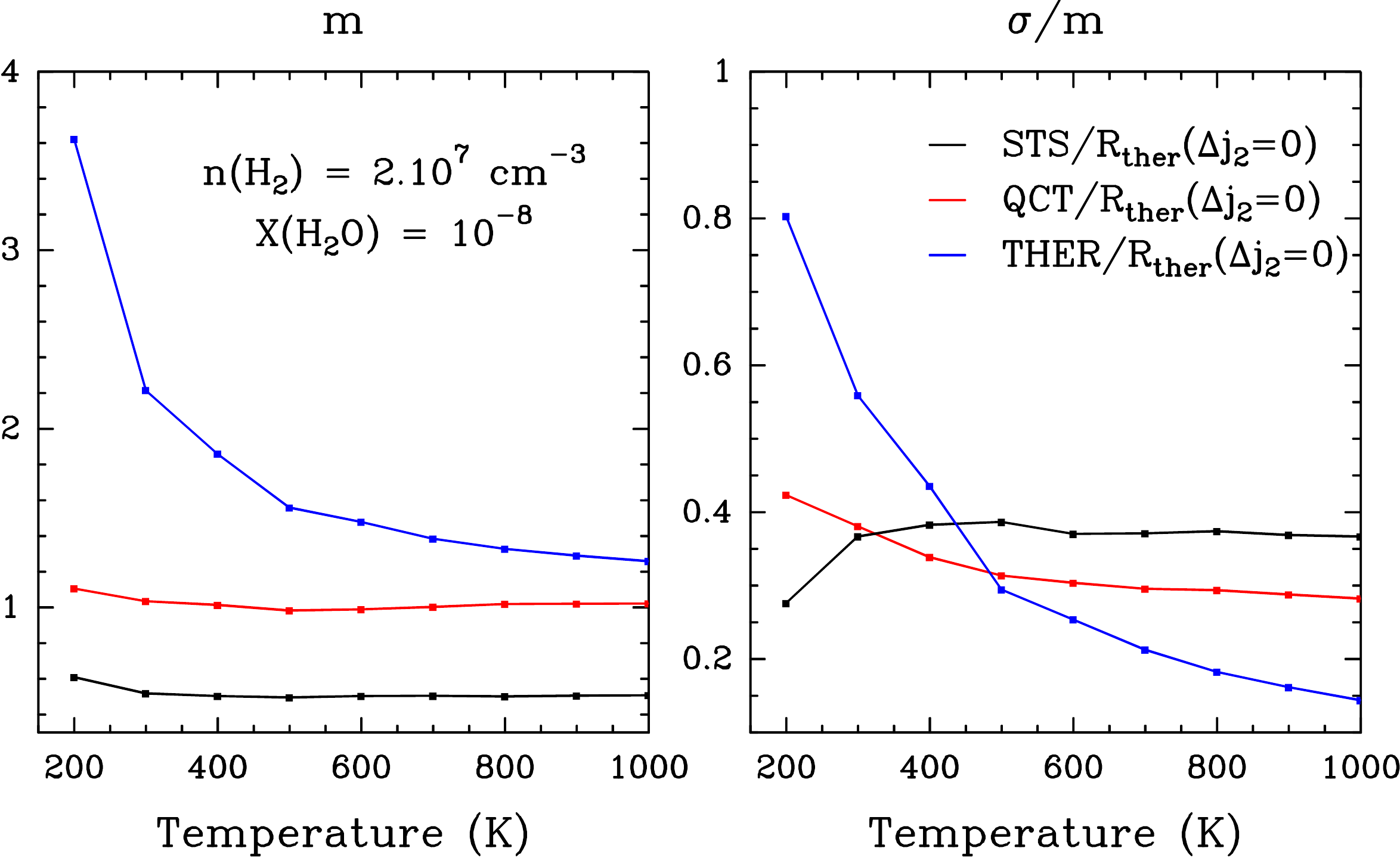}
\end{center}
\caption{Comparison of the mean (left column) and normalized standard deviation (right column) obtained for the ratios $\bar{I}/\bar{I}^{approx}$ and dealing with p--H$_2$.
The approximation consists in neglecting the terms $C_{ij}(j_2 \to j_2')$ with $\Delta j_2 \geq 2$ in the calculation 
of the thermalized CRC. The results obtained with this approximation are compared to the 
quantum state--to--state CRC (black lines) from \citet{dubernet2009,daniel2011}, the QCT CRC (red lines) from \citet{faure2007} and thermalized CRC (blue lines) from \citet{dubernet2009,daniel2011}.}
\label{comp31}
\end{figure}

A striking feature, when considering the comparison of the STS and thermalized CRC 
arises from the behaviour at low temperature. Indeed, the highest differences are found in the range 200-400 K between
those sets. Simply considering the fact that the population of the p--H$_2$ $j_2 = 2$ level increases with 
temperature, one would expect the differences between those two sets to behave similarly, 
to reach a maximum when the fundamental level is depopulated. On the contrary, the highest 
differences are found around 200 K and tend to diminish while the temperature increases. 
This effect is due to the terms $C_{ij}(j_2 \to j_2' )$ with $\Delta j_2 \geq 2$. In order to quantify the influence
of those terms, we used an ad--hoc set of CRC. In this set, the thermalized CRC are calculated setting to 0
all the CRC with $\Delta j_2 \neq 0$. The comparison between the results based on the STS, QCT and thermalized
CRC with this ad--hoc set is shown in Fig. \ref{comp31}. The comparison is made at a density $n$(H$_2$) = $2.10^7$ cm$^{-3}$
and for a water abundance $X$(H$_2$O) = 10$^{-8}$. The choice of these parameters is based on 
the results shown in Fig. \ref{comp10} where it can bee seen that for both the QCT and STS CRC, these parameters 
correspond  to the maximum differences encountered. 
Considering the mean value obtained for the ratio $\bar{I}^{Ther}/\bar{I}^{approx.}$ we see that the mean value has its maximum 
at 200 K and then decreases with temperature. This proves that the terms $C_{ij}(j_2 \to j_2' )$ with $\Delta j_2 \geq 2$ 
are responsible of the high differences encountered at low temperatures. Considering the mean value for the ratios 
$\bar{I}^{QCT}/\bar{I}^{approx.}$, we find that it has a constant value $\sim 1.0$ over the whole temperature range.
The normalized standard deviation decreases from 0.4 to 0.3 as the temperature increases. In other words, the ad-hoc set of CRC
and the QCT CRC give the same results within $\sim 30\%$. This shows that the main drawback of the QCT approximation
is that it does not correctly reproduce the terms $C_{ij}(0 \to 2)$. This result is not surprising since these transitions have large rate coefficients when there is a quasi--resonance between the H$_2$O and H$_2$ rotational levels, i.e. transitions with 
$\Delta_{ij}>500$ K. At the QCT level, quasi--resonant effects are not included properly owing to the approximate quantization procedure. \\
 \\

 \subsection{Discussion}

In the previous sections, we compared the results of various CRC sets 
for o--H$_2$O. The comparisons were made
by considering the lines that involve energy levels below the 20$^{th}$ one 
(i.e. $E_u < 900 K$). With respect
to the water symmetry, calculations were also performed for the p--H$_2$O molecule 
and the conclusions were found to be similar, i.e. the quantum thermalized and QCT CRC 
 give qualitatively similar results for the line 
intensities, when considering both o--H$_2$ and p--H$_2$ as collisional partners, 
with maximum differences for the bulk of the lines of the order of a factor 3.

\begin{figure}[h!]
\begin{center}
\includegraphics[angle=270,scale=0.35]{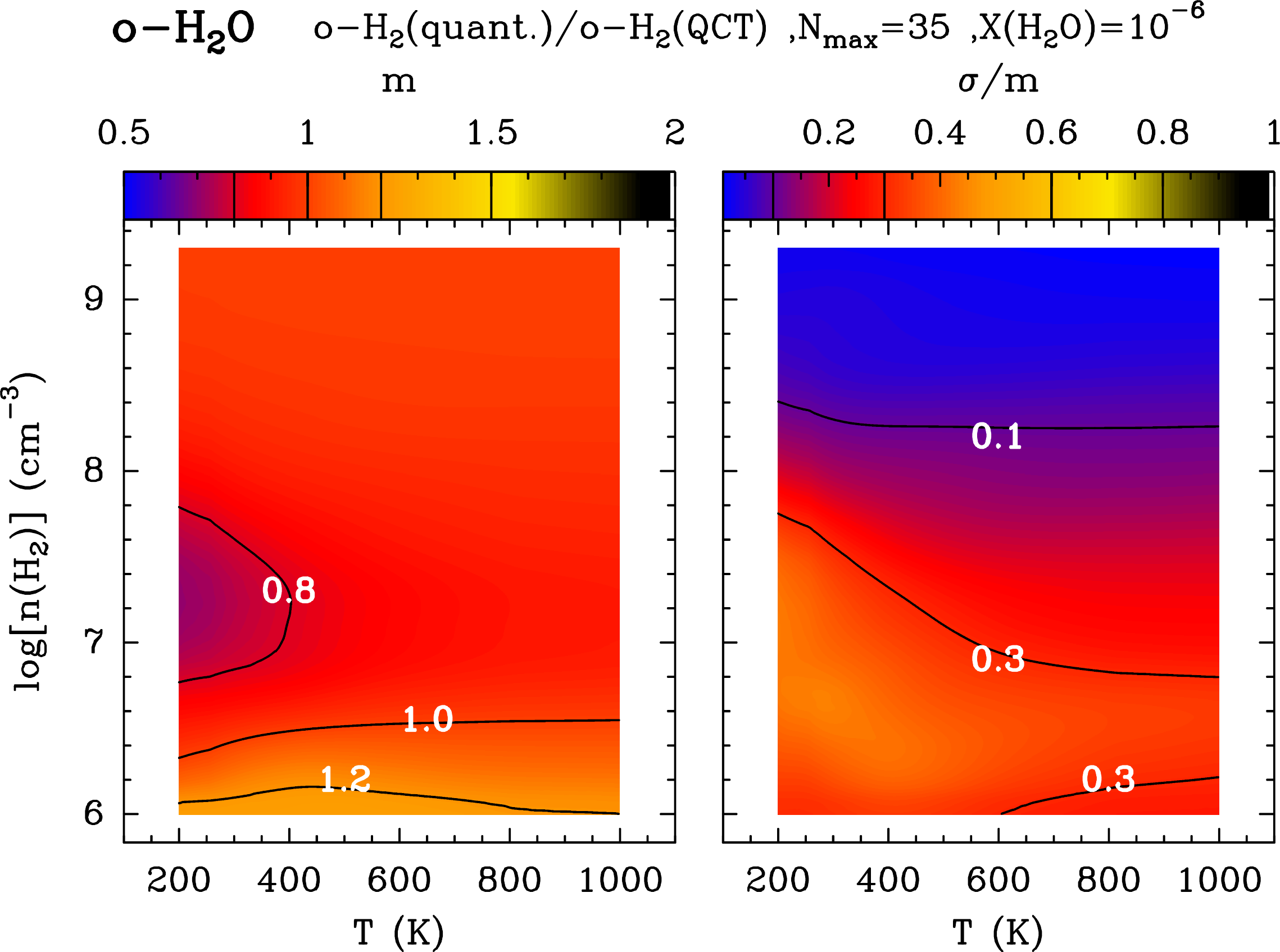}
\end{center}
\caption{Mean value and normalised standard deviation obtained comparing the quantum CRC from \citet{dubernet2009,daniel2011}
and QCT CRC from \citet{faure2007},
for a water abundance of $\chi$(H$_2$O) = $10^{-6}$ cm$^{-3}$. The lines retained in the comparison have an energy 
below the one of the 35$^{th}$ level of o--H$_2$O.}
\label{levels}
\end{figure}

With respect to the number of H$_2$O levels considered, a similar statistical analysis has been performed
considering the lines that involve the first 35$^{th}$ energy levels. The differences for the ratios were
found to be larger for the lines with $E_u > 900$ K. This qualitatively results in larger 
variations of the normalised standard deviations (i.e. would magnify the scale of $\sigma$ on 
Fig. \ref{comp1}, \ref{comp2} and \ref{comp3}). 
This is illustrated on Fig. \ref{levels} where the mean value and standard deviations are
given, for the ratio between the o-H$_2$ STS and QCT CRC. In this figure,
we consider all the lines with energy level below the 35$^{th}$ level and for the case 
of a water abundance of $\chi$(H$_2$O) = 10$^{-6}$. The results are to be compared 
to the results shown on Fig. \ref{comp3} where only the first 20$^{th}$ levels were considered. 

Finally, to a first order, it can be seen that the variations of intensities from one set to 
the other are linearly correlated to the CRC variations. This is illustrated by considering the 
critical densities related to the various sets of CRC discussed in the previous sections which are reported 
in Table \ref{table2}. Comparing, as an example, the critical densities related to the p--H$_2$ QCT and 
thermalized CRC, we can see that the maximum differences is of a factor 3, which is 
similar to the maximum difference found for the intensity ratio (see Sect. \ref{thermalized}).

\begin{table*}[h]
\begin{center}
\begin{tabular}{|c|c|c|ccc|ccc|}

\hline
transition & $\nu$ (GHz)  & He & \multicolumn{3}{c}{p--H$_2$} &  \multicolumn{3}{c|}{o--H$_2$}  \\ \hline
               &                       &       & QCT & THER & STS &   QCT & THER & STS  \\ 
\hline
     6 $_{   1,  6}$ -  5 $_{   2,  3}$ &       22. &      1.2(3) &      1.1(3) &      3.3(2) &      4.2(2) &      1.1(3) &      2.9(2) &      2.9(2) \\ 
     4 $_{   1,  4}$ -  3 $_{   2,  1}$ &      380. &      5.2(6) &      2.9(6) &      2.8(6) &      4.9(6) &      1.5(6) &      1.4(6) &      1.4(6) \\ 
     4 $_{   2,  3}$ -  3 $_{   3,  0}$ &      448. &      5.9(6) &      3.9(6) &      3.5(6) &      6.1(6) &      1.4(6) &      1.5(6) &      1.5(6) \\ 
     1 $_{   1,  0}$ -  1 $_{   0,  1}$ &      557. &      8.1(7) &      4.8(7) &      3.4(7) &      6.9(7) &      3.1(7) &      1.4(7) &      1.4(7) \\ 
     3 $_{   1,  2}$ -  3 $_{   0,  3}$ &     1097. &      6.7(8) &      1.4(8) &      2.7(8) &      5.8(8) &      1.1(8) &      1.1(8) &      1.1(8) \\ 
     3 $_{   1,  2}$ -  2 $_{   2,  1}$ &     1153. &      2.2(8) &      1.1(8) &      1.1(8) &      2.0(8) &      4.9(7) &      4.3(7) &      4.3(7) \\ 
     3 $_{   2,  1}$ -  3 $_{   1,  2}$ &     1163. &      9.1(8) &      1.5(9) &      3.5(8) &      7.2(8) &      7.4(8) &      1.5(8) &      1.5(8) \\ 
     5 $_{   2,  3}$ -  5 $_{   1,  4}$ &     1411. &      1.6(9) &      1.1(9) &      6.6(8) &      1.6(9) &      4.3(8) &      2.6(8) &      2.6(8) \\ 
     2 $_{   2,  1}$ -  2 $_{   1,  2}$ &     1661. &      1.9(9) &      2.6(9) &      9.7(8) &      1.9(9) &      1.4(9) &      4.2(8) &      4.2(8) \\ 
     2 $_{   1,  2}$ -  1 $_{   0,  1}$ &     1670. &      1.3(9) &      8.4(8) &      6.6(8) &      1.1(9) &      6.7(8) &      3.3(8) &      3.3(8) \\ 
     4 $_{   3,  2}$ -  5 $_{   0,  5}$ &     1714. &      2.2(8) &      2.2(8) &      1.4(8) &      2.0(8) &      2.2(8) &      8.7(7) &      8.7(7) \\ 
     3 $_{   0,  3}$ -  2 $_{   1,  2}$ &     1717. &      1.8(9) &      6.3(8) &      8.6(8) &      1.4(9) &      4.6(8) &      4.1(8) &      4.1(8) \\ 
     5 $_{   2,  3}$ -  4 $_{   3,  2}$ &     1919. &      1.8(9) &      7.1(8) &      8.0(8) &      1.1(9) &      3.0(8) &      4.8(8) &      4.8(8) \\ 
     3 $_{   3,  0}$ -  3 $_{   2,  1}$ &     2196. &      4.6(9) &      2.0(9) &      2.5(9) &      4.0(9) &      1.0(9) &      1.2(9) &      1.2(9) \\ 
     5 $_{   1,  4}$ -  5 $_{   0,  5}$ &     2222. &      6.4(9) &      5.9(8) &      3.1(9) &      6.1(9) &      4.2(8) &      1.2(9) &      1.2(9) \\ 
     4 $_{   2,  3}$ -  4 $_{   1,  4}$ &     2264. &      7.1(9) &      5.4(9) &      3.3(9) &      5.8(9) &      1.9(9) &      1.5(9) &      1.5(9) \\ 
     4 $_{   3,  2}$ -  4 $_{   2,  3}$ &     2463. &      1.0(10) &      3.0(9) &      4.7(9) &      6.9(9) &      2.2(9) &      2.5(9) &      2.5(9) \\ 
     4 $_{   1,  4}$ -  3 $_{   0,  3}$ &     2641. &      9.9(9) &      4.2(9) &      4.9(9) &      6.5(9) &      3.0(9) &      3.0(9) &      3.0(9) \\ 
     2 $_{   2,  1}$ -  1 $_{   1,  0}$ &     2774. &      1.1(10) &      5.3(9) &      5.0(9) &      7.0(9) &      4.2(9) &      2.8(9) &      2.8(9) \\ 
     5 $_{   1,  4}$ -  4 $_{   2,  3}$ &     2971. &      1.6(10) &      1.9(10) &      7.4(9) &      8.8(9) &      7.5(9) &      6.0(9) &      6.0(9) \\ 
     5 $_{   0,  5}$ -  4 $_{   1,  4}$ &     3013. &      1.7(10) &      8.3(9) &      8.7(9) &      1.1(10) &      5.5(9) &      6.0(9) &      6.0(9) \\ 
     6 $_{   1,  6}$ -  5 $_{   0,  5}$ &     3655. &      3.9(10) &      1.9(10) &      2.0(10) &      2.2(10) &      1.2(10) &      1.5(10) &      1.5(10) \\ 
     4 $_{   2,  3}$ -  3 $_{   1,  2}$ &     3807. &      4.6(10) &      3.6(10) &      2.4(10) &      2.9(10) &      2.1(10) &      1.7(10) &      1.7(10) \\ 
     3 $_{   2,  1}$ -  2 $_{   1,  2}$ &     3977. &      4.3(10) &      4.4(10) &      2.1(10) &      3.2(10) &      3.9(10) &      1.2(10) &      1.2(10) \\ 
     3 $_{   3,  0}$ -  3 $_{   0,  3}$ &     4457. &      1.3(10) &      1.3(10) &      9.7(9) &      1.7(10) &      6.7(9) &      5.3(9) &      5.3(9) \\ 
     3 $_{   3,  0}$ -  2 $_{   2,  1}$ &     4512. &      1.0(11) &      4.6(10) &      4.4(10) &      5.7(10) &      3.1(10) &      3.3(10) &      3.3(10) \\ 
     4 $_{   3,  2}$ -  3 $_{   2,  1}$ &     5107. &      2.0(11) &      9.7(10) &      8.6(10) &      1.1(11) &      5.5(10) &      7.4(10) &      7.4(10) \\ 
     5 $_{   2,  3}$ -  4 $_{   1,  4}$ &     6646. &      4.9(11) &      4.9(11) &      1.3(11) &      3.4(11) &      8.8(10) &      1.2(11) &      1.2(11) \\ 
     4 $_{   3,  2}$ -  3 $_{   0,  3}$ &     7368. &      5.1(11) &      5.1(11) &      8.2(10) &      4.8(11) &      8.5(10) &      7.5(10) &      7.6(10) \\  \hline
\end{tabular}
\end{center}
\caption{Critical densities for the lines that involve the first 15$^{th}$ o--H$_2$O energy levels, at T = 200 K. 
The critical densities are given for the He CRC of \citet{green1993}, the QCT CRC of \citet{faure2007} and the quantum CRC
of \citet{dubernet2009,daniel2011}. In the latter case, the values are given for both STS and thermalized CRC.}
\label{table2}
\end{table*}%

\section{The o--H$_2$ / p--H$_2$ dichotomy} \label{OTPR}

To date, apart from H$_2$O, only a few collisional systems have been treated considering both
o--H$_2$ and p--H$_2$ as collisional partners 
(i.e. CO by \citet{wernli2006}, HC$_3$N by \citet{wernli2007}, SiS by \citet{lique2008,klos2008}, H$_2$CO by \citet{troscompt2009a}, 
HNC by \citet{dumouchel2011}, CN$^-$ by \citet{klos2011},  SO$_2$ by \citet{cernicharo2011}
,HF by \citet{guillon2012}, HDO by \cite{faure2012}). 
Except in the CN$^{-}$ case, 
a common conclusion is that the o--H$_2$ CRC are larger than for p--H$_2$.
In the particular case of water, the differences are rather large, with CRC
for o--H$_2$ that can be larger by up to a factor 10 compared to p--H$_2$. 
This implies that the population of the high energy levels through collisions with o--H$_2$ will
be favoured.
To date, only few detailed discussions concerning the differences in the excitation
of a given molecule and considering the effects introduced by the differing
collisional partners exist. 
A short discussion of the influence of the H$_2$ OTPR ratio was made 
by \citet{cernicharo2009}, where 
a few water vapour lines were considered for the case of a model describing a protoplanetary disk.
Such a discussion has also been done for the H$_2$CO molecule. \citet{troscompt2009} found that the 
H$_2$ OTPR was crucial in explaining the 
excitation of the $1_{10}-1_{11}$ line, and that it is possible to accurately
constrain the H$_2$ OTPR ratio from its observation. On the other hand,
\citet{guzman2011} found that for the physical conditions which are typical of 
the Horsehead nebula, the H$_2$CO lines observed in their study are insensitive 
to the H$_2$ OTPR.
A similar conclusion was obtained by \citet{parise2011} for the deuterated isotopomers of 
H$_3^+$ which were found to be marginally affected by the H$_2$ OTPR for the conditions
typical of prestellar cores (see also \citet{pagani2009}).
In what follows, we discuss some characteristics of the water vapour
excitation with respect to collisions with o--H$_2$ or p--H$_2$.

In order to study the influence of the H$_2$ symmetry, we compare
the ratio of the values taken by $\bar{I}$ for
the quantum CRC (i.e. $\bar{I}^{ortho}/\bar{I}^{para}$).
In the comparison, we consider both the STS and thermalized
CRC. 
For a given line, we compute the mean value 
and the normalised standard deviation by summing over all the the models that
correspond to the parameter space defined in Sec. 2.
We use the same selection for the lines as previously done and which are given by 
the criteria indicated in Sec. 2. 

\subsection{State--to--State rate coefficients}

\begin{figure}[h]
\begin{center}
\includegraphics[angle=0,scale=0.5]{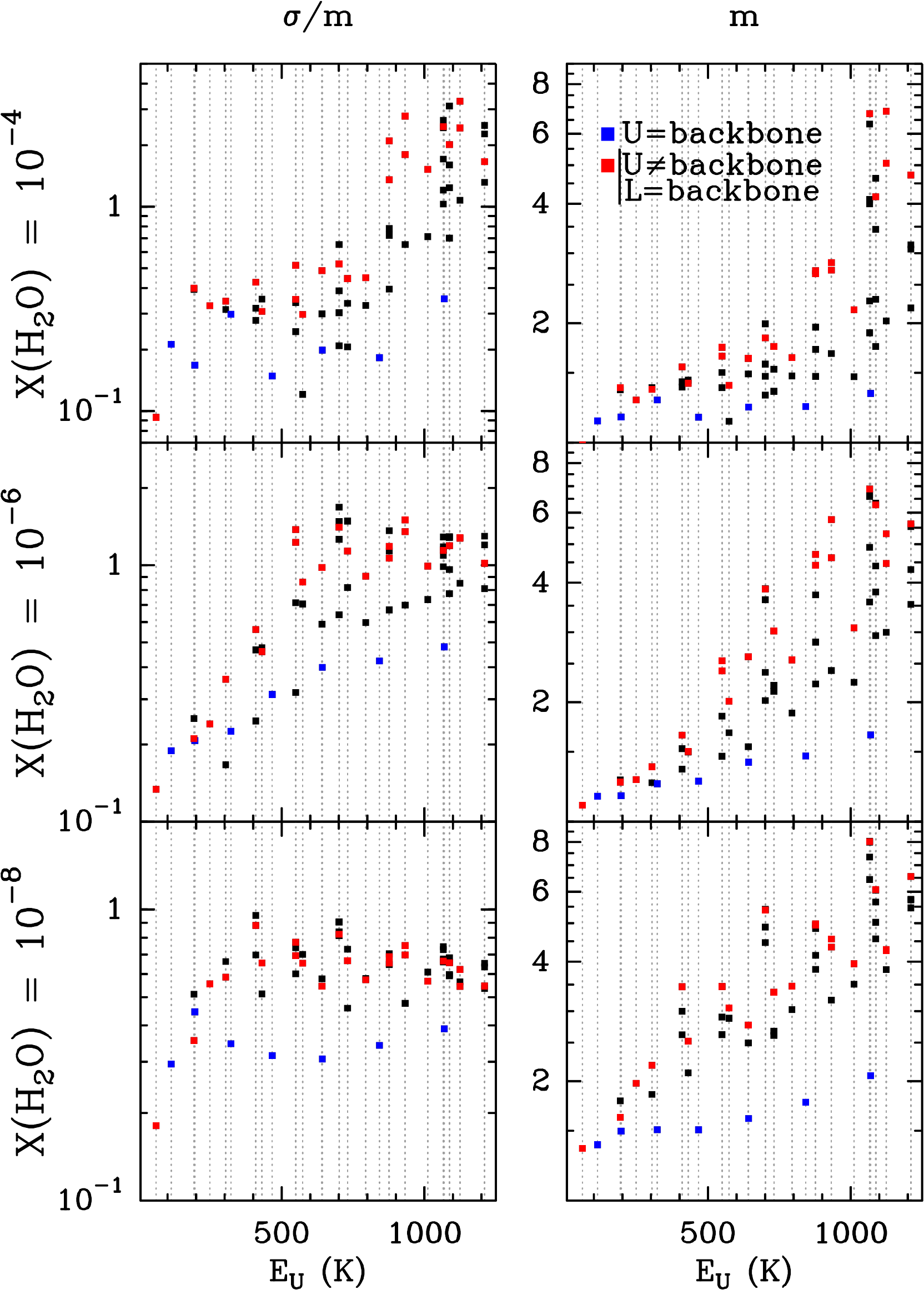}
\end{center}
\caption{Mean value (left column) and normalised standard deviation (right column) 
for the $\bar{I}^{ortho}/\bar{I}^{para}$ ratios as a function of the energy of the upper level
of the radiative transitions. The intensities are obtained with the STS CRC \citet{dubernet2009,daniel2011}. 
The transitions 
that involve an upper energy level which is backbone are indicated by blue points. The transitions 
where the upper level is not a backbone level while the lower level is backbone are indicated by red points.}
\label{OTP}
\end{figure}

The mean value and normalised standard deviation are plotted in 
Fig. \ref{OTP}, for the three value of the water abundance considered in this work. 
In this figure, we see that the mean value and normalised standard deviation will depend
differently on the collisions with o--H$_2$ or p--H$_2$ according to the 
energy of the upper level involved in the transition. Additionally, there is a correlation
between the sensitivity of the transition to the o--H$_2$/ p--H$_2$ symmetry 
with the position of the upper energy level on the J--ladder\footnote{As a reminder, 
H$_2$O is an asymmetric top with quantum numbers noted either 
$J, K_a,K_c$ or $J,K_+,K_-$. A backbone level corresponds to the lower level in energy for a given 
value of the principal quantum number $J$. This level corresponds to the lowest possible value for the difference
$K_a-K_c$.}.
Qualitatively, an increase of the energy of the upper level will be accompanied by an increase
of both the mean value and normalised standard deviation. For the transitions that involve an upper energy level
below 500 K, we find that independently of the water abundance, the mean value is around 1
and the normalised standard deviation takes low values (below 1). Those lines are thus marginally
affected by the collisional partner. For the lines with an 
upper energy level above 500 K, the transitions which are the less affected 
by the nature of the collisional partner are the transitions 
where the upper energy level is a backbone level (blue points in Fig. \ref{OTP}).
For these transitions, the mean value remains relatively low (typically below 2.5)
for all the values of the water abundance. The normalised standard deviation is minimal for those transitions, 
especially when the water abundance is low. 
This implies that the intensity of 
these transitions will be only marginally affected by the symmetry of the collisional partner.
The transitions which are the most affected are the one 
for which the upper level is not a backbone level while the lower level is backbone level (red points in Fig. \ref{OTP}).
These transitions should be good indicators of the H$_2$ OTPR of the gas.

\subsection{Thermalized rate coefficients}

In the previous section, on the base of the quantum state--to--state CRC, 
it was shown that the H$_2$ OTPR would influence differentially the intensities
of the water transitions. In this section, the same analysis is performed using the thermalized quantum CRC,
for high gas temperatures ($T_k > 200$ K) and low gas temperatures ($T_k < 100$ K). 

\subsubsection{High temperature}

\begin{figure}[h!]
\begin{center}
\includegraphics[angle=0,scale=0.5]{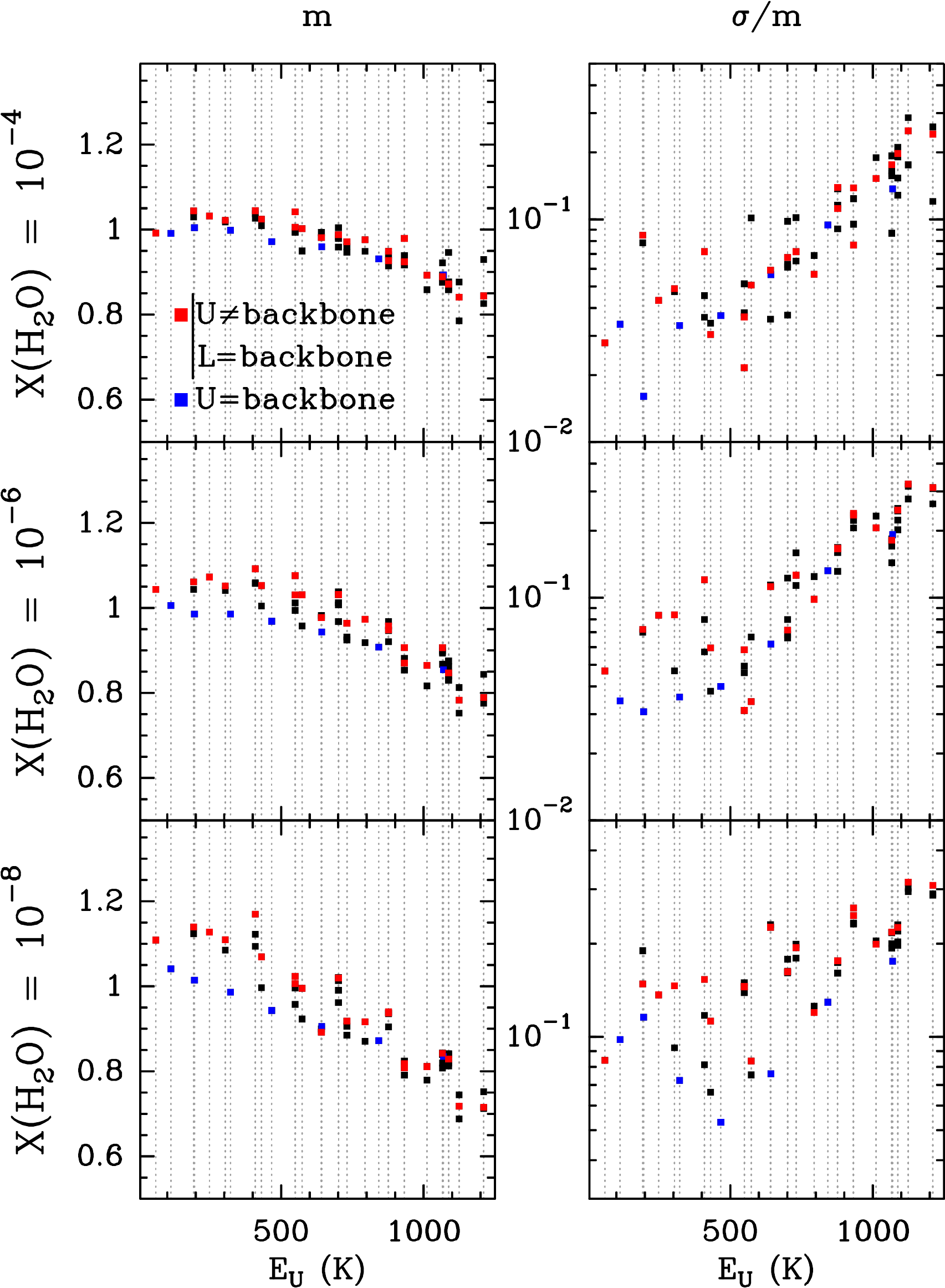}
\end{center}
\caption{Same as Fig. \ref{OTP} but considering the quantum thermalized CRC from \citet{dubernet2009,daniel2011}.}
\label{OTP-2}
\end{figure}
Figure \ref{OTP-2} shows the means and normalised standard deviations for the 
ratios $\bar{I}^{ortho}/\bar{I}^{para}$ obtained using the thermalized quantum CRC. From this 
figure, it appears that the bulk of the lines are mainly insensitive to the H$_2$ OTPR, 
contrary to what was obtained using the STS CRC. For the transitions that involve 
an upper energy level with energy below 500 K, the mean value is found to be close to 1,
independently of the water abundance considered.
For the transitions that involve higher levels, we find that there is a departure from the 
mean value of 1 but the departure is modest, since most of the intensity ratios are in 
the range $0.7 < m < 1.1$.
Additionally, the normalised standard deviations are low, i.e.  $\sigma/m < 0.3 $, which means that $\sim 70\%$
of the lines considered in the analysis show variations of less than 30\% around the mean value. 
Interestingly, the transitions that involve energy levels higher than 500 K are found to be globally 
brighter when considering the collisions with p--H$_2$ than when considering the collisions with o--H$_2$.
This effect is induced by the increase of magnitude of the CRC with $\Delta E_{ij} > 500$ K 
which has been discussed in Sec. 2.

\subsubsection{Low temperatures}

\begin{figure*}[h!]
\begin{center}
  %\subfigure[]{\includegraphics[angle=270,scale=0.35]{./GRAPHS/LOWT_110-101.pdf}} \quad
  %\subfigure[]{\includegraphics[angle=270,scale=0.35]{./GRAPHS/LOWT_212-101.pdf}} \quad
    %\subfigure[]{\includegraphics[angle=270,scale=0.35]{./GRAPHS/LOWT_221-110.pdf}} \\
    %\subfigure[]{\includegraphics[angle=270,scale=0.35]{./GRAPHS/LOWT_221-212.pdf}} \quad
  %\subfigure[]{\includegraphics[angle=270,scale=0.35]{./GRAPHS/LOWT_303-212.pdf}} \quad
    %\subfigure[]{\includegraphics[angle=270,scale=0.35]{./GRAPHS/LOWT_312-303.pdf}} \\
  %\subfigure[]{\includegraphics[angle=270,scale=0.35]{./GRAPHS/LOWT_321-212.pdf}} \quad
    %\subfigure[]{\includegraphics[angle=270,scale=0.35]{./GRAPHS/LOWT_330-221.pdf}} \quad
  %\subfigure[]{\includegraphics[angle=270,scale=0.35]{./GRAPHS/LOWT_414-303.pdf}} \\
     \includegraphics[angle=0,scale=1.0]{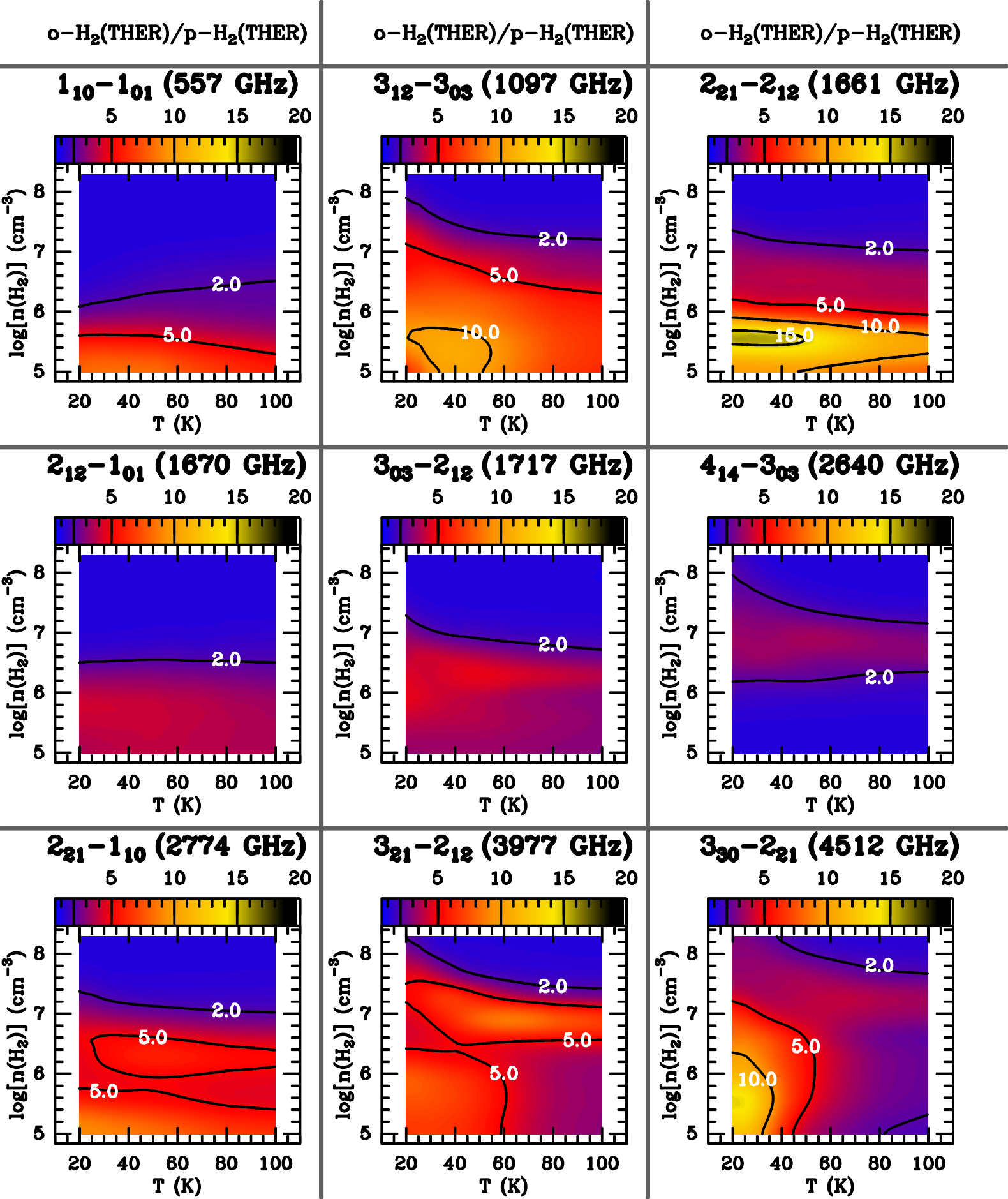}
\end{center}
\caption{Comparison of the ratio $\bar{I}^{ortho}/\bar{I}^{para}$ for a few o--H$_2$O transitions that 
involve its lowest energy levels. The results are derived from the thermalised CRC from \citet{dubernet2009,daniel2011} and are given for the temperature range T$_K$ = 20--100 K.}
\label{comp11}
\end{figure*}

To discuss the low temperature regime, we carried out models with free parameters in the range : 
T $\in$ [20K;100K], n(H$_2$) $\in$ [10$^5$ ; 2 $\, 10^8$] cm$^{-3}$ and 
$\chi$(H$_2$O) $\in$ $\left\{ 10^{-8} ; 10^{-6} ; 10^{-4} \right\}$. 
The mean values and normalised standard deviations are calculated for each line by considering 
all the models of the grid. In Table \ref{table1}, we give the mean values and standard 
deviations for all the lines. From this table, it appears that all the lines 
which are considered are affected by the H$_2$ OTPR.
The main effect, as discussed earlier, is to obtain an increase of intensity when considering o--H$_2$
as a collisional partner, for the transitions with upper energy level below 500 K. On the other hand, for the levels
with an upper energy level above 500 K, considering p--H$_2$ as a collisional partner 
results in higher intensities.
Additionally, the line intensities are affected differentially by the symmetry of the collisional partner. 
This differential effect is presented in Fig. \ref{comp11} where the intensity ratios are
shown for some of the lines that involve the lowest energy levels of o--H$_2$O. 
From this figure it can be seen that under specific physical conditions, the intensity ratio can take high values
for certain lines (i.e. larger than 15, like for example for the $2_{21}-2_{12}$) while it remains low for other lines 
(i.e. below 2, like for example for the $4_{14}-3_{03}$).
Since, the lines depend differentially on the H$_2$ OTPR, it is in principle possible to determine the H$_2$ 
OTPR for the molecules of the gas from an accurate modelling of the water 
line intensities observations. In practice, however, this can be a difficult task due to the dependence 
of the line intensities on other parameters of the model, like the gas temperature,
H$_2$ volume density and geometry of the source. Moreover, this is only feasible if 
it relies on a large set of observations. This puts strong limits on the usefulness of water to 
derive the H$_2$ OTPR at low temperature (T $<$ 50 K) since only a few transitions 
will be observable with a reasonable sensitivity (i.e. with RMS $<$ 10 mK).

\begin{table*}
\begin{center}
\begin{tabular}{|cccc|cc|cc|cc|}
\hline
transition & $\nu$ (GHz) & $\lambda$ ($\mu$m) & E$_u$ (K) & \multicolumn{2}{c}{$\chi$(H$_2$O) = $10^{-8}$} &  \multicolumn{2}{|c|}{X(H$_2$O) = $10^{-6}$} & \multicolumn{2}{c|}{X(H$_2$O) = $10^{-4}$}\\ 
\hline
\multicolumn{4}{|c|}{} & mean & $\sigma/m$ & mean & $\sigma/m$ & mean & $\sigma/m$ \\ \hline
  1$_{  1,  0}$ -  1$_{  0,  1}$ &      557. &       538. &       61. &          2.8 &          0.8 &          1.4 &          0.4 &          1.3 &          0.3 \\
  3$_{  1,  2}$ -  3$_{  0,  3}$ &     1097. &       273. &      249. &          3.2 &          0.7 &          2.7 &          0.8 &          2.3 &          1.1 \\
  3$_{  2,  1}$ -  3$_{  1,  2}$ &     1163. &       258. &      305. &          2.8 &          0.7 &          2.6 &          0.8 &          2.1 &          0.8 \\
  2$_{  2,  1}$ -  2$_{  1,  2}$ &     1661. &       180. &      194. &          2.5 &          0.6 &          2.1 &          0.6 &          1.5 &          0.3 \\
  2$_{  1,  2}$ -  1$_{  0,  1}$ &     1670. &       180. &      114. &          2.1 &          0.6 &          1.7 &          0.5 &          1.6 &          0.5 \\
  3$_{  0,  3}$ -  2$_{  1,  2}$ &     1717. &       175. &      197. &          2.2 &          0.5 &          2.0 &          0.7 &          1.2 &          0.4 \\
  3$_{  3,  0}$ -  3$_{  2,  1}$ &     2196. &       136. &      411. &          2.6 &          0.7 &          2.0 &          0.6 &          1.8 &          0.6 \\
  4$_{  2,  3}$ -  4$_{  1,  4}$ &     2264. &       132. &      432. &          2.9 &          0.6 &          2.3 &          0.8 &          2.0 &          0.8 \\
  4$_{  1,  4}$ -  3$_{  0,  3}$ &     2640. &       114. &      323. &          1.8 &          0.4 &          1.7 &          0.6 &          1.5 &          0.4 \\
  2$_{  2,  1}$ -  1$_{  1,  0}$ &     2774. &       108. &      194. &          2.5 &          0.7 &          2.5 &          0.9 &          2.0 &          0.5 \\
  5$_{  1,  4}$ -  4$_{  2,  3}$ &     2971. &       101. &      575. &          1.9 &          0.4 &          1.6 &          0.4 &          1.3 &          0.6 \\
  5$_{  0,  5}$ -  4$_{  1,  4}$ &     3013. &        99. &      468. &          1.7 &          0.3 &          1.6 &          0.5 &          1.6 &          0.6 \\
  6$_{  1,  6}$ -  5$_{  0,  5}$ &     3655. &        82. &      644. &          1.1 &          0.2 &          1.1 &          0.2 &          1.1 &          0.3 \\
  4$_{  2,  3}$ -  3$_{  1,  2}$ &     3807. &        79. &      432. &          2.4 &          0.4 &          2.0 &          0.7 &          1.8 &          0.6 \\
  3$_{  2,  1}$ -  2$_{  1,  2}$ &     3977. &        75. &      305. &          3.2 &          0.8 &          3.4 &          1.6 &          2.4 &          1.1 \\
  7$_{  0,  7}$ -  6$_{  1,  6}$ &     4167. &        72. &      843. &          0.5 &          0.2 &          0.7 &          0.2 &          0.8 &          0.1 \\
  3$_{  3,  0}$ -  3$_{  0,  3}$ &     4457. &        67. &      411. &          2.6 &          0.5 &          2.5 &          0.8 &          2.3 &          0.7 \\
  3$_{  3,  0}$ -  2$_{  2,  1}$ &     4512. &        66. &      411. &          2.4 &          0.4 &          2.1 &          0.7 &          1.9 &          0.6 \\
  6$_{  2,  5}$ -  5$_{  1,  4}$ &     4600. &        65. &      796. &          0.8 &          0.2 &          1.0 &          0.3 &          1.0 &          0.3 \\
  4$_{  3,  2}$ -  3$_{  2,  1}$ &     5107. &        59. &      550. &          1.2 &          0.4 &          1.3 &          0.3 &          1.0 &          0.3 \\
  4$_{  4,  1}$ -  3$_{  3,  0}$ &     6076. &        49. &      702. &          0.5 &          0.4 &          0.7 &          0.3 &          0.7 &          0.4 \\
  5$_{  3,  2}$ -  4$_{  2,  3}$ &     6249. &        48. &      732. &          0.5 &          0.6 &          0.8 &          0.3 &          0.8 &          0.3 \\
  5$_{  2,  3}$ -  4$_{  1,  4}$ &     6646. &        45. &      642. &          0.9 &          0.4 &          1.1 &          0.3 &          1.0 &          0.4 \\
  4$_{  3,  2}$ -  3$_{  0,  3}$ &     7368. &        41. &      550. &          1.7 &          0.2 &          1.5 &          0.4 &          1.3 &          0.2 \\
\hline
\end{tabular}
\caption{Mean values and normalised standard deviations for the ratios  $\bar{I}^{ortho}/\bar{I}^{para}$ obtained
with the thermalized CRC from \citet{dubernet2009,daniel2011}. 
The parameters correspond to gas temperatures in the range $20 < T_k < 100$ K and
$ 10^5 <$ $n$(H$_2$) $<$ $2.10^8$ cm$^{-3}$.}
\label{table1}
\end{center}
\end{table*}

\section{Dust radiative pumping} \label{discussion}

\begin{figure}[h!]
\begin{center}
\includegraphics[angle=270,scale=0.35]{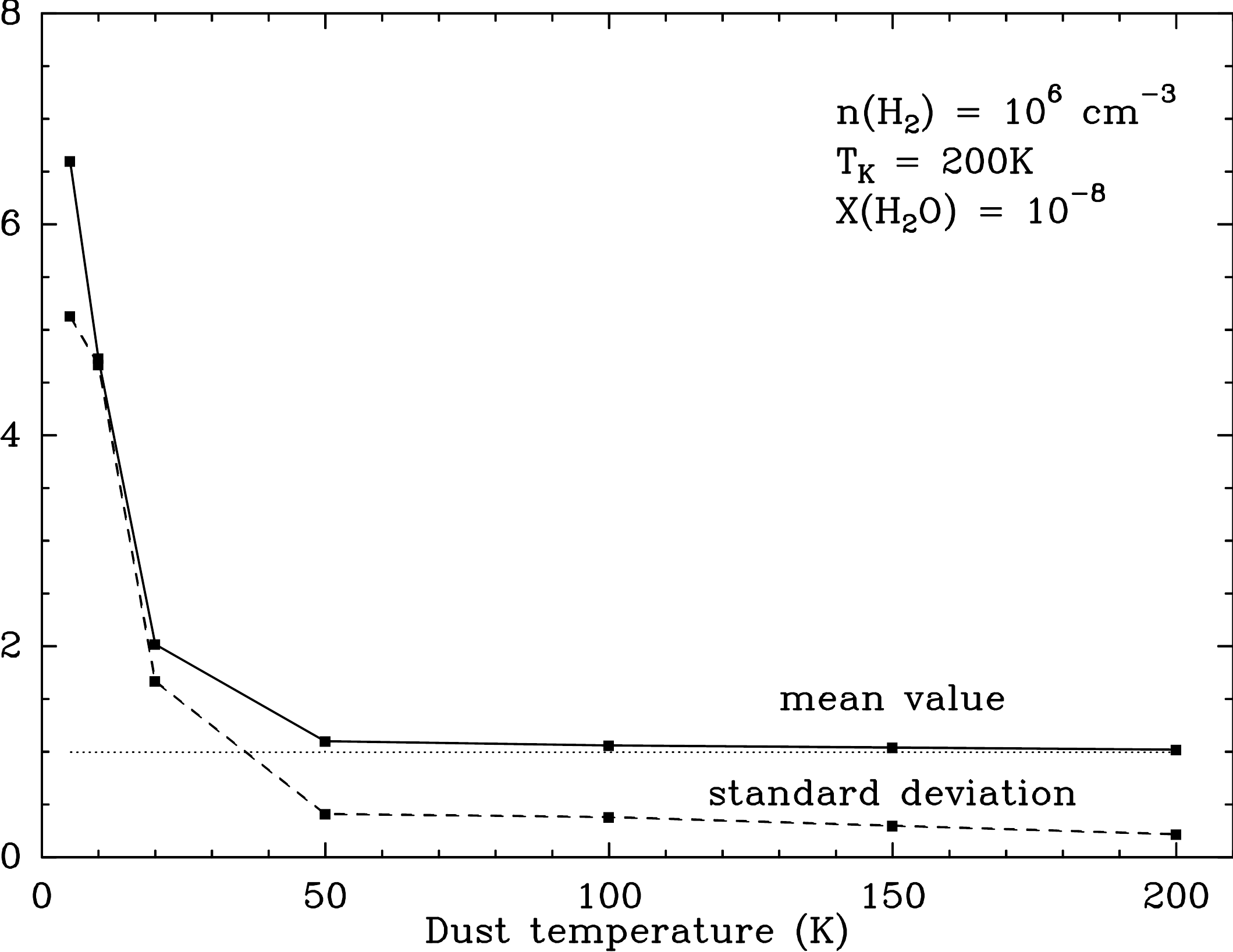}
\end{center}
\caption{Mean value and standard deviation as a function of the dust temperature.
The ratio considered are obtained considering the quantum ortho-- and para-- STS CRC from \citet{dubernet2009,daniel2011}.}
\label{dust}
\end{figure}

All the comparisons performed in Sec. 2 and 3 were done ignoring the possibility of 
pumping by infrared and sub--millimetre dust radiation.
This would correspond to an extreme case, not physically 
relevant to all astrophysical objects.
Indeed, as an example, radiative pumping by continuum photons plays an important role in the H$_2$O
excitation in AGB circumstellar envelopes.
In short, considering this additional mechanism in the population
of the water energy levels would reduce the differences between line intensities
obtained from differing CRC. This is illustrated in Fig \ref{dust}, where we consider
the $\bar{I}^{ortho}/\bar{I}^{para}$ ratios obtained from the STS CRC.
In this example, we choose to compare the results based on these CRC
since it has been found in Sec. \ref{OTPR} that the respective intensities show 
large differences and thus enable to emphasize the role played by dust radiation.
The model parameters are: $n$(H$_2$) = 10$^6$ cm$^{-3}$, T$_K$ = 200 K and $\chi$(H$_2$O) = 10$^{-8}$. 
In the model, we assume a gas--to--dust mass ratio of 100 and the dust composition corresponds 
to a mixture of astrophysical silicates and amorphous carbon grains, with opacities taken from
\cite{draine1984}. In the modelling, we varied the dust temperature (T$_d$) from 5 K to 200 K. 
Moreover, to compute $\bar{I}$, we assume for the background temperature : $T_{bg} = T_{d}$.
From this figure,
we see that while the mean values and standard deviations are high at low dust temperatures ($m \sim 7$ and
$\sigma \sim 5$), they are considerably reduced when T$_d$ increases. For dust temperatures above 50 K,
we obtain $m \sim 1$ and $\sigma < 0.2$, which implies that the influence of differing CRC starts
to be minimal since the population of the H$_2$O energy levels is dominated by 
radiative pumping and not by the collisions anymore.

\section{Conclusions} \label{conclusions}

We performed non--local non--LTE excitation and radiative transfer calculations aiming at comparing the line intensities
predicted for water vapour when making use of differing collisional rate coefficients sets. The collisional rate coefficients 
sets compared are the He quantum rate coefficients \citep{green1993}, 
H$_2$ quantum rate coefficients of \citep{dubernet2006,dubernet2009,daniel2010,daniel2011} and H$_2$ 
quasi--classical rate coefficients \citep{faure2007}. The comparison was performed at relatively high 
temperature (200 K $<$ T$_K$ $<$ 1000 K) and an emphasis was made on the comparison of the 
H$_2$ collisional rate coefficient sets, since the quantum rate coefficients were only made lately available
and many recent astrophysical studies made use of the QCT calculations. In the absence of radiative 
pumping by dust photons, it was found
that the results based on the quantum and QCT rate coefficients sets will lead to line intensities
that qualitatively agree, i.e. which are of the same order of magnitude,
 for the parameter space considered in this work. However, 
the H$_2$O line emission can differ by a factor of the order of $\sim$ 3, in the regime of 
low water abundance ($\chi$(H$_2$O) $\sim$ 10$^{-8}$) and moderate H$_2$ volume densities 
($n$(H$_2$) $<$ 10$^7$ cm$^{-3}$).

These differences should not drastically affect the conclusions 
obtained from the modelling of the water excitation in astrophysical objects but
the water vapour abundance derived on the base of the QCT rate coefficients will differ 
with the one derived from the quantum rate coefficients, with differences still up 
to a factor of $\sim$ 3. We note, however, that such differences will be attenuated by the presence of a 
dust continuum source of radiation. Additionally, masing lines are not considered 
in the present analysis. A similar comparison for the case of masing lines is 
presented in \citet{Daniel12} where the impact induced by the CRC on the lines that
will be observable with ALMA is discussed. Finally, we emphasize on the fact that 
the impact of the various CRC sets is discussed on the base of a statistics over the most
intense lines. The current results can thus be affected to some extent by the choice of the subset 
of lines used in the analysis.

The differences found between the QCT and quantum CRC can give clues on the 
uncertainty which is introduced in the modeling due to the uncertainties on the rate coefficients.
Indeed, as it was discussed in \citet{dubernet2009} and \citet{daniel2010}, the QCT and quantum CRC
typically agree within a factor of 3 for what concerns the highest rate coefficients. In the current study,
we obtain the same factor between the line intensities obtained with two sets. 
Therefore, in a first approximation, 
the error made on the rate coefficients will translate similarly on line intensities. Recently, 
\citet{yang2011} found a good agreement between experimental integral cross sections and quantum
calculations, showing the good accuracy of the PES on which are based the quantum calculations.
Depending on the energy of the level considered and the gas temperature, it was said in \citet{dubernet2009} 
and \citet{daniel2010} that the accuracy of the quantum CRC will range from a few \% to a few ten \% and such
errors should scale linearly on line intensities.

We performed additional test calculations with an ad--hoc set of thermalised collisional rate coefficients in which
the state--to--state rate coefficients associated to the H$_2$ transitions with $\Delta j_2 \neq 0$ were removed.
The comparison of the results obtained with this ad--hoc set and the QCT rate coefficients showed
a particularly good agreement. Over the temperature range T = 200--1000 K, the intensities predicted 
with those two sets agree within 30\%. It is concluded that the main drawback of the QCT approximation
is that it does not correctly reproduce the terms $C_{ij}(0 \to 2)$.

By considering the quantum state--to--state and quantum thermalized collisional rate coefficients, it 
was found that line intensities will be largely affected by the consideration of the first 
excited state, in the case of p--H$_2$. The differences start to be non--negligible (i.e. greater than 20\%) 
for temperatures higher than $T_K \sim 60$ K, reaching a maximum of a factor $\sim$ 3 around 200 K. 
On the other hand, the intensities predicted with the state--to--state and thermalized collisional rate coefficients
are similar for the collisions that involve o--H$_2$. 
Such a behaviour can be extrapolated to other molecules that show strong differences 
for the collisions between o--H$_2$ and p--H$_2$. Indeed, such differences arise because of
the interaction with the quadrupole of H$_2$ and significant differences for the collisional rate coefficients with
p--H$_2$ or o--H$_2$ imply that the state--to--state rate coefficients for p--H$_2$ in $j_2 = 0$ or $j_2 = 2$
will show large differences too.
Therefore, the use of state--to--state rather than thermalized
collisional rate coefficients for such molecules will lead to an overestimate of the molecular abundances
for temperatures higher than T$_K$ $\sim$ 60 K.

By comparing the line intensities obtained using as a collisional partner 
either o--H$_2$ or p--H$_2$, we found that under particular 
physical conditions ($n$(H$_2$) and T$_K$), the water lines will be differentially affected
by the symmetry of the H$_2$ molecule. This effect is obtained when the gas temperature is low, 
i.e. $T_K$ $<$ 100 K. On the contrary, at high temperature, the lines become insensitive 
to the H$_2$ symmetry. Since some transitions will show large intensity 
variations with respect to the H$_2$ symmetry, the modelling of the 
water excitation may provide a mean to derive the H$_2$ ortho--to--para ratio. The lines
to be considered in such an estimate will however depend on the physical conditions
prevailing in the object under study and it is necessary to perform a 
case--by--case modelling.

\begin{acknowledgements}
The authors want to thank the anonymous referee whose detailed comments 
enabled to improve the content of the current article.
This paper was partially supported
within the programme CONSOLIDER INGENIO 2010, under grant ''Molecular
Astrophysics: The Herschel and ALMA Era.- ASTROMOL'' (Ref.: CSD2009-
00038). We also thank the Spanish MICINN for funding support through
grants AYA2006-14876 and AYA2009-07304. JRG is supported by a Ram\'on y Cajal
research contract from the spanish MICINN and co-financed by the European Social Fund.
\end{acknowledgements}

\end{document}